\documentclass[10pt]{article}
\pdfoutput=1

\usepackage{etoolbox}
\patchcmd{\thebibliography}{\leftmargin\labelwidth}
    {\itemsep 0pt \leftmargin\labelwidth}{}{}

\usepackage[paper=a4paper,dvips,top=1.5cm,left=1.5cm,right=1.5cm,
    foot=1cm,bottom=1.5cm]{geometry}
    
\usepackage{bbm} 

\usepackage{graphicx}
\usepackage{mathptmx}
\usepackage{braket}
\usepackage{multicol}
\usepackage{multirow}
\usepackage[table]{xcolor}
\usepackage{lscape}
\usepackage{amssymb}
\usepackage{amsthm}
\usepackage{flushend}
\usepackage[numbers]{natbib}
\usepackage{ctable}
\usepackage{makecell}
\usepackage{array}
\newcolumntype{L}[1]{>{\raggedright\let\newline\\\arraybackslash\hspace{0pt}}m{#1}}
\newcolumntype{C}[1]{>{\centering\let\newline\\\arraybackslash\hspace{0pt}}m{#1}}
\newcolumntype{R}[1]{>{\raggedleft\let\newline\\\arraybackslash\hspace{0pt}}m{#1}}

\usepackage{hyperref}

\title{Cross-border Portfolio Investment Networks\\ and Indicators for Financial Crises}

\author{
        \textsc{$\mbox{Andreas C. Joseph}^1$}\thanks{\href{mailto:andreas.ch.joseph@gmail.com}{\texttt{andreas.ch.joseph@gmail.com}}}
            \qquad
        \textsc{$\mbox{Stephan E. Joseph}^2$}
            \qquad
        \textsc{$\mbox{Guanrong Chen}^1$}
        \mbox{}\\
         \small{$\mbox{ }^1$ Centre for Chaos and Complex Networks, Department of Electronic Engineering} \\ 
         \small{City University of Hong Kong, Hong Kong S.A.R., China} \\
         \small{$\mbox{ }^2$ Faculty of Economics and Business, Universitat de Barcelona, Catalonia, Spain}
}

\begin{document}
%
%
\maketitle
\textbf{Abstract} - Cross-border equity and long-term debt securities portfolio investment networks are analysed from 2002 to 2012, covering the 2008 global financial crisis. They serve as network-proxies for measuring the robustness of the global financial system and the interdependence of financial markets, respectively. Two early-warning indicators for financial crises are identified: First, the algebraic connectivity of the equity securities network, as a measure for structural robustness, drops close to zero already in 2005, while there is an over-representation of high-degree off-shore financial centres among the countries most-related to this observation, suggesting an investigation of such nodes with respect to the structural stability of the global financial system. Second, using a phenomenological model, the edge density of the debt securities network is found to describe, and even forecast, the proliferation of several over-the-counter-traded financial derivatives, most prominently credit default swaps, enabling one to detect potentially dangerous levels of market interdependence and systemic risk.\\[.2cm]
\tableofcontents
\clearpage
\normalsize
\section{Introduction}
\label{sec:intro}
Since the 2008 global financial crisis (GFC'08), there has been growing awareness of the interdependence of international financial markets and the \textit{systemic risk}, i.e.\ the probability of a system-wide failure, resulting from their intrinsic entanglement \cite{net_opp}. It was shown how even supposedly minor distress of individual actors in a network of major financial institutions could lead to the annihilation of large economic value \cite{debt_rank}. Conventional macroeconomic models, such as dynamic stochastic general equilibrium models \cite{GSEM_2, GSEM_3, econ_model_rep}, which assume the existence of a stable equilibrium with disturbing shocks coming from outside the system, failed to predict, or even describe, the GFC'08.\\
According to the current understanding \cite{net_opp,comp_der,cds_net,cds_risk,crotty,db_cds}, the proliferation of certain financial products prior to the GFC'08, such as collateralised mortgage obligations and credit default swaps (CDS), has led to a global network of strong interdependence of financial institutions with its geographical center in the United States (US). The crisis was then triggered by a relatively small shock inside the US mortgage market, which spread globally over this network \cite{inter_bank}, leading to a dry-up of inter-bank lending. Subsequently, this resulted in the actual crisis, putting many systematically important institutions at risk \cite{debt_rank}. While it is estimated that the US banking sector suffered losses of about 1.8\ trillion USD \cite{crotty}, the value of global financial assets declined by around 16\ trillion USD \cite{mgi}, not taking into account knock-on effects generated by this decline. Thus, the GFC'08, originating from a destabilising internal shock, which propagated over a complex network of strong interactions. This scenario is exactly in the realm of network science.\\
The study of large-scale economic networks has seen considerable progress in recent years, as demonstrated by \cite{corp_contr,contr_pow,uni_sys_risk,boot_net,contr_pow_2}. However, much of the data relevant to the detailed investigation of global inter-institutional networks are deemed strategically important for the involved institutions, and are not available. On the other hand, actual macroscopic dependencies among actors in international financial markets, such as the involvement of whole economies and the macroscopic interplay between different sectors of the global financial system may be probed by means of \textit{proxy networks}, using aggregated flows/positions on the inter-economy level. \\
Motivated by this observation, we investigate two types of cross-border portfolio investment networks (PIN), namely equity- (E) and long-term debt (LD) securities, which represent major components of international capital markets. The respective networks are labelled LD-PIN and E-PIN, where nodes are given by individual countries and directed and weighted edges by consolidated investment positions, measured in USD, originating from residents/institutions in one country to residents/institutions in another \cite{cpis}. Details about data and the used methodologies can be found in the sections \hyperref[sec:meth]{Methodology} and \hyperref[sec:dis]{Discussion}. The main figures and table are given in \hyperref{si:fig}[Appendix\ 1] and \hyperref{si:fig}[Appendix\ 2], respectively.  A presentation of the relevant concepts of graph theory, as well as additional information on both networks and results, are provided in the \hyperref{si:si}[Appendix\ 3: Supplementary Information].\\
The GFC'08 is clearly reflected in both PIN, where an overall reduction in investment positions (contraction) is observed, followed by a re-bounce as early as 2010. Our analysis reveals two early-warning indicators for potential financial crises. On the one hand, the algebraic connectivity of the E-PIN, interpreted as an indicator for the structural robustness of the global financial system and, as such, for the world-economy, drops sharply as early as 2005. This observation is associated with the emergence of a sparsely connected group of countries, particularly involving Middle Eastern countries, the United Kingdom (UK) and several high-degree off-shore financial centres (OFC), where we detect a general over-representation of OFC among countries associated with this structural instability. On the other hand, the edge density of the LD-PIN, as a network-proxy to measure the interdependence of financial markets, scales with the total market values of several over-the-counter-traded (OTC-traded) financial derivative products, which have been directly linked to the crisis, such as CDS \cite{oecd_der,der_over,crotty}, but also equity-linked derivatives (ELD). Based on this, a simple phenomenological model is introduced, which allows for the description of the proliferation of such derivative products and, as such, for the detection of potentially high levels of market interdependency. We propose a dynamic monitoring mechanism which, taking the GFC'08 as a testing ground, generates clear warning signals between 6-12\ months ahead of the crisis.\\
\section{Results}
\subsection{General Properties of PIN and the GFC'08}
\label{sec:results-1}
Both PIN are rapidly growing in terms of the numbers of nodes $N$ and edges $M$ in the years before the GFC'08, as can be seen in \hyperref[fig:1]{Fig.\ 1}, where all quantities are shown with respect to their year-2002 values (dotted reference line). A summary of network statistics in given in \hyperref[si:1]{App.\ 3: S-1}. For both the E- and the LD-PIN, basic network parameters scale with major macroeconomic quantities. Taking into account the very different time resolutions of data points, the number of edges in the E-PIN (a) tracks quite well global stock market indices, such as the $S\&P\;\mathrm{Global}\,1200$ \cite{sp,econ_stats}. The initial mismatch can be explained through the previous burst of the``dot-com bubble'' and the following mild recession, which affected mainly the US \cite{dot_com}. In addition, the total trade volume between PIN countries scales roughly with the volume of the E-PIN, as shown in \hyperref[si:5]{App.\ 3: S-5}. Since stock markets, which the E-PIN is inherently connected to, and international trade are widely accepted measures for economic performance, the E-PIN offers a network-proxy for measuring the state of the world economy.\\
For the LD-PIN, the temporal evolution of the edge density $\rho^{LD}_p$ at the minimal percolation edge threshold (see \hyperref[sec:meth]{section Methodology}) mirrors qualitatively the evolution of gross-market value (GMV) of OTC-traded ELD, as well as the notional outstanding amount (NOA) of CDS \cite{bis_der}, which is shown in panel (b) of\hyperref[fig:1]{Fig.\ 1}. \\
The prices/values of financial derivatives are linked to an underlying asset, such as a stock, debt security or commodity, which by definition creates market interdependences, while their primarily uses are risk management (hedging) and speculation \cite{oecd_der,der_over,crotty,imf_der}. The value of ELD is derived from the price of some stock or a stock index. As such, ELD provide an indirect link between the two PIN, where it is believed that cross-asset hedges and capital structure arbitrage trades are the most important drivers \cite{jpm_cross,der_over}. Surprisingly, the inter-connectedness between equity and debt markets created in this way has so far not attracted any attention, but might complement the understanding of the GFC'08. One reason for this could be the large variety of ELD, such as single- or multi-stock and index forwards, options and swaps, where the macroscopic impact of individual products is difficult to assess.\\
Debt securities can be traded before the borrowed amount (principal) is repaid, i.e.\ before their maturity, or be used as base assets for financial derivative products such as CDS. These are credit derivative products, where the credit exposure from one or several third parties is passed from the buyer to the seller. This activity is expected to lead to additional inter-connectedness between a larger number of debtors and creditors. The general understanding is that CDS played an important role in transmitting the shock from the 2007-US subprime mortgage crisis, through large insurers, to the international banking system \cite{comp_der,cds_net,cds_risk,crotty,db_cds,oecd_der}. These observations make the LD-PIN a suitable network-proxy to measure the inter-connectedness and, consequently, the interdependence of financial markets, which have been identified as major factors contributing to financial crises \cite{preis}.\\
Note that NOA and GMV of financial derivative products are interpreted as two different risk measures. NOA describes the market value with respect to the underlying base assets (face value), such as mortgages, bonds or stocks. It may be taken as a measure for the overall market interconnectedness/interdependence, while it does not represent the actual amount at risk. GMV is the cost to replace existing contracts at the current market rate, which may be less coupled to the underlying assets than to the prevailing economic environment. It is seen to be a better (short-term) risk indicator than NOA.\\
\hyperref[fig:2]{Figure\ 2} illustrates structural properties of both PIN in terms of their cumulative distributions of node strength (a, b) and eigenvector centrality $C_{ev}$ (c, d). These two measures evaluate a country's importance within both networks from different points of view. Node strength, which is the sum of all in-coming and out-going investment positions, measures the size of a country in a network, while eigenvector centrality is a recursive measure, which accounts for a country's embedding into the weighted and directed network topology (see \hyperref[si:1]{App.\ 3: S-1} for more information on both measures).\\
An interesting feature of both PIN is the strong hierarchical structure, as depicted by their node strength distributions. These can be classified as being ``super-heavy-tailed'', in the sense that node strengths span several orders of magnitude, while there is an $\mathcal{O}\left(1\right)$-probability for most values.\\
We see that the E-PIN is strongly dominated by the US and to a minor degree by the UK during all times, which is especially true for the eigenvector centrality (see \hyperref[fig:2]{Fig.\ 2-c}). Its wave-like structure indicates a multi-layered topology with the US at the center. Taken together, such a configuration renders the E-PIN fragile against shocks originating from its most central nodes \cite{net_attack}.\\
Both measures are more homogeneously distributed in the LD-PIN (see \hyperref[fig:2]{Figs.\ 2-b and -d}), where the most central nodes are Japan, France, the US and Germany. The proliferation of ELD, as shown in \hyperref[fig:1]{Fig.\ 1-b}, is now interpreted as an additional (geographical) shift of weight from the LD-PIN to the E-PIN, and towards the US. This can be understood as follows. Given the observed relation between ELD and the edge density of the LD-PIN, a high centrality of countries in Europe and Asia in the LD-PIN links them to the E-PIN, where the US is the dominating country. This configuration is expected to amplify shocks, as for example those from the US subprime mortgage crises of 2007.\\
Next, we will investigate how the GFC'08 is reflected in both PIN. PIN are large-scale economic structures. The total volume contained in all PIN together, which includes short-term debt (SD) securities, is of the same order in magnitude as world-GDP (gross-domestic product, \cite{wb_data}; see blue line in \hyperref[fig:3]{Fig.\ 3-a}), peaking at approximately $56\,\%$ of world-GDP at the beginning of 2008. The crisis leads to an overall contraction of both networks, with the total volume reduced to less than $45\%$ of world-GDP, and a partly re-bounce to pre-crisis levels as early as 2010.\\
Both effects are stronger in the E-PIN, which experiences a faster growth and larger contraction, than the LD-PIN before and during the crisis. About $80\,\%$ of the total contraction during the crisis result from the E-PIN, which shrinks by about $47\,\%$ (or $10\,$trillion\ USD) from 2008-2009. Note, at this point, that the contraction of both PIN already captures a large part of the decrease in value of global financial assets resulting from the GFC'08 \cite{mgi}.\\
The higher variability of the E-PIN during the GFC'08 is attributed to the \textit{volatile} nature of its links as compared to those of the LD-PIN. This is due to the fact that investment in equity is generally riskier than investment in long-term debt. Equity securities can be readily sold in a crisis, additionally experiencing dramatic changes in value (here, edge weight). This is not the case for long-term debt securities, where links are by definition more durable with fixed weights, since future returns on investment are generally determined at the time of issuance. Moreover, there has been a reduction of liquidity in the debt markets \cite{debt_market}, as a consequence of the crisis, which prevented the LD-PIN from contracting substantially.\\
Besides an overall expansion of both networks, we observe a gradual shift from debt markets to equity markets in the years before the GFC'08 \cite{mgi}, which is reflected in a changing composition of the total PIN, as seen in \hyperref[fig:3]{Fig.\ 3-b}, where the temporal evolution of the respective fractions of volume of the total PIN (blue line in \hyperref[fig:3]{Fig.\ 3-a}) contained in the E- and LD-PIN are shown. This is interpreted as a growing fragility of the global financial system because the E-PIN is more susceptible to financial crises than the LD-PIN. We remark that the strong anti-correlation between E- and LD-PIN volumes is expected because they represent the majority of cross-border portfolio investment, as reported in \cite{cpis}.\\
A simple network measure for the structural robustness of the E-PIN against edge or node failures (removals) is the algebraic connectivity $\lambda_1^E$ (see \hyperref[si:1]{App.\ 3: S-1} for a detailed description), where a zero value means the decomposition of the E-PIN into two disconnected components. $\lambda_1^E$ drops sharply in the beginning of 2005 and reaches an all-time low in the beginning of 2007 (\hyperref[fig:3]{Fig.\ 3-c}), pointing to a structural fragility of the E-PIN prior to the GFC'08. This is an intriguing fact from the network perspective, because both PIN are dense with edge densities $\rho>0.3$, average in-/out-degrees well above $20$ and a minimal edge weight of $e_{th}=52$\ million USD during all times (see \hyperref[si:3]{App.\ 3: S-3}).\\
We have seen that the build-up of the GFC'08 can be well described under the PIN framework, confirming the conventional understanding, yet providing new insights from the network point of view. Two rather simple early-warning indicators for potential financial crises emerge from this analysis: The \textit{algebraic connectivity} $\lambda_1^E$ of the E-PIN, as a measure for structural robustness, and the \textit{edge density} $\rho_{p}^{LD}$ of the LD-PIN, as a measure for the level of interdependence within the global financial system.
\subsection{Indicator I: The Algebraic Connectivity of the E-PIN}
\label{sec:results-2.1}
A (hypothetical) decomposition of the E-PIN is expected to lead to a global economic crisis, since it disconnects international equity markets. The precise value of $\lambda_1^E$ depends on the numbers of nodes and edges and the network topology, rendering values for different times not directly comparable to each other. However, as explained in \hyperref[si:1]{App.\ 3: S-1}, for most practical applications with $N\gg 2$, \textit{one} is an effective upper bound and, as such, the interval $]0,\,1]$ offers a quasi-absolute scale. An alternative scale is set by the value of the next-largest eigenvalue of the normalised Laplacian $\lambda_2\geq \lambda_1$, where one can consider the difference of both values $\Delta\lambda=\lambda_2-\lambda_1$, given that $\lambda_2$ is approximately constant and smaller than one. Looking at \hyperref[fig:3]{Fig.\ 3-c}, where $\lambda_2^E$ and $\Delta\lambda^E$ are given for reference, both scales point to an increased structural fragility of the E-PIN before and during the GFC'08, which we will investigate more closely now. Two fundamental questions here concern the properties of this instability, namely if it affects the whole network and to what extend, and which countries are most-related to it.\\
To address the first issue, we consider the graph bi-section, given by the ordering of nodes according to the signatures of the entries of the eigenvector $\vec{v}_1^E$ associated with $\lambda_1^E$ (Fiedler vector), which we call Fiedler bi-section \cite{com_dec_1,GT_book2}. As stated in \hyperref[si:1]{App.\ 3: S-1}, this is expected to result in a good graph bi-section, in terms of a low cut ratio $R(s_{+},s_{-})=s_{cut}/(|s_+|\cdot|s_-|)$, when $\lambda_1^E$ is close to zero. \\
To evaluate the quality of Fiedler bi-sections of both PIN, we compare them with different types of random bi-sections, where the nodes of a PIN are randomly divided into two groups of equal size, random sizes of $1$ to $N-1$ nodes or sizes of the corresponding Fiedler bi-section (labelled Fiedler-like). The results of this exercise are summarised in \hyperref[si:3]{App.\ 3: S-3}, where we use the \textit{cut depth} $D_{cut}$ of a bi-section of weighted graphs, which relates the average weight of edges between the two partitions of a bi-section to the average edge weight of the whole network. A value larger than one indicates that two partitions are connected by stronger-than-average edges, while the opposite is true for a value smaller than one. We see that Fiedler bi-sections, which are far outside the given $99\,\%$-confidence intervals, are significantly different from any random bi-section of the E- as well as the LD-PIN. \\
Thus, given the moderately low values of $\lambda_1^E$ observed in the years before the GFC'08, the Fiedler vector can be used to detect the fault lines in the E-PIN. The details of this bi-section are summarised in \hyperref[si:7]{App.\ 3: S-7}, where the nodes contained in the smaller sections $s_{\mathrm{small}}^{\lambda,E}$ and their in- and out-degrees (numbers in parentheses), as well as the corresponding fractions of nodes $f_{\mathrm{small}}^{\lambda,E}$ for all years, are given. Note that such a bi-section is always possible, while one expects an approximately equal partition of a graph for large values of $\lambda_1$, since this will, on average, minimise the cut ratio in the absence of a particular $\lambda_1$-instability. This is exactly what was observed for the E-PIN at the beginning and the end of the observation period and for the LD-PIN at all times.\\
We use the parameter triple $T(E)=(\lambda_1^E,\,f_{\mathrm{small}}^{\lambda,E},\,D_{cut}^{\lambda,E})$ for the description of the observed structural instability of the E-PIN prior and during the GFC'08. Considering \hyperref[fig:3]{Fig.\ 3-d}, where the percentage of nodes $f_{\mathrm{small}}^{\lambda,E}$ is given by the red numbers, one sees that the drop of $\lambda_1^E$ at the beginning of 2005 can be explained by the emergence of a sparsely connected group of countries, which is expressed through the small cut depth $D_{cut}^{\lambda,E}$ at that time. \\
This picture is changing dramatically during the GFC'08, when the cut depth rises substantially in the beginning of 2009, while $\lambda_1^E$ recovers partially. This is explained trough a very peculiar network configuration, where the US and UK, which are the most central nodes in the E-PIN (see \hyperref[fig:2]{Fig.\ 2-c}), enter the much smaller section $s_{\mathrm{small}}^{\lambda,E}$ (see \hyperref[si:7]{App.\ 3: S-7}), indicating considerable fault lines within the E-PIN during the crises. \\
Next, we identify the nodes which can be related strongest to the emergence of the fragility of the E-PIN at the beginning of 2005. Many of the high-degree nodes in $s_{\mathrm{small}}^{\lambda,E}$ during 2005-2008 are classified as OFC, defined as ``\textit{a country or jurisdiction that provides financial services to non-residents on a scale that is incommensurate with the size and the financing of its domestic economy}'' (\cite{def_ofc}, page 9), such as the Bahamas, the Cayman Islands or Guernsey. On average, OFC make up around one third of nodes in the E-PIN, while lists of OFC according to \cite{def_ofc} and found in the E-PIN are given in \hyperref[si:2]{App.\ 3: S-2}. This finding raises the question if a tight integration of OFC into the E-PIN can be associated with the observed structural instability.\\
To answer this question, we remove multiple groups of $1$-$10$ nodes and observe the resulting changes in $\lambda_1^E$ for 2005-2008. The new E-PIN is defined as the largest strongly-connected component after that node removal. It is reasonable to assume that a node or a group of nodes are relevant for the formation of the observed $\lambda_1^E$-instability if removing it leads to a considerable increase in $\lambda_1^E$ to a value larger than $0.5$: either a clear jump of $\lambda_1^E$ to a value around $0.7$, or the same qualitative picture as shown in \hyperref[fig:3]{Fig.\ 3-c}, is observed. Such instability lifting between 2005-2008 can be partial or complete, i.e.\ during some years or the whole period.\\
The only two countries, which completely lift $\lambda_1^E$ when being removed, are Bahrain and Kuwait. This suggests that the observed structural instability is centred around these two Middle Eastern countries, which are both \textit{not} highly central to the E-PIN in terms of their total degrees or strengths. One may therefore conclude that the observed $\lambda_1^E$-instability is not of systemic relevance. Note, however, that this is a static picture, where it is not at all clear how a full or partial break-up of a graph, caused by the removal of some edges or nodes, affects the \textit{dynamics} in the network, where scenarios such as cascading failures are imaginable. Furthermore, the modern financial system is a global one and the situation turns out to be more complex when considering groups of countries, as is expected from the network perspective.\\
For pairs of countries, Bermuda and Guernsey are the only combination which can partially lift $\lambda_1^E$, bar pairs involving Bahrain or Kuwait. For triples, we found two groups made of the UK, Jersey, and either Bermuda or Egypt, which partially lift the instability; again, bar those involving previously found single countries or pairs. For larger groups, the picture turns out to be more complex. Note that an exhaustive check of all possible combinations of groups of sizes $n=4$-$10$ is computationally infeasible because of the super-exponentially growing number of possible groups. We therefore use a statistical \textit{two-step breadth-first} search algorithm to detect combinations of countries that are involved in the formation of the $\lambda_1^E$-instability. After removing Bahrain and Kuwait from the pool of possible selections, we perform the following computations:\\
\begin{enumerate}
\item \textit{Breadth:} For $n\in \lbrace 4,\,\dots,\,10\rbrace$, randomly draw $10^4$ samples and separately calculate $\lambda_1^E$ between 2005 and 2008. Next, order all countries according to the frequency that they have been involved in groups lifting $\lambda_1^E$ for at least one year.
\item \textit{Depth:} For $n\in \lbrace 4,\,\dots,\,10\rbrace$, draw again $10^4$ samples and check the $\lambda_1^E$-instability, but now half (round-up) of nodes of each sample is drawn from the corresponding ten most-frequent countries from step one, and the other half is drawn from the remaining pool. This procedure is based on the observation that approximately ten most-frequent countries show a relative frequency in their numbers of occurrences higher than what one expects from a pure random selection. Again, order all countries according to their frequencies of being found in groups lifting $\lambda_1^E$ for at least one year.
\end{enumerate}
Since this search routine is non-deterministic for $n=4$-$10$, we perform five full rounds and average the results over all outcomes. The final results are summarised in \hyperref[tab:1]{Tab.\ 1}, where the average probability of finding a combination of nodes, whose removal leads to a jump in  $\lambda_1^E$, is given by $p_{\lambda}$. Its step-like behaviour for rising values of $n$ is caused by the rounding rule of step two. We should point out that there are no considerable fluctuations in the overall search results between different rounds, which is especially true for large $n$, such as $n=9$ and $n=10$, with mean-over-standard-deviation ratios of $p_{\lambda}$ of $0.017$ and $0.008$, respectively.\\
The countries mostly associated with the $\lambda_1^E$-instability are shown in the right column. For $n>3$, the $15$ most-frequently found countries are ordered according to their frequencies of occurrence. There is a group of OFC (Bermuda, Guernsey and Jersey), the UK and Egypt, which are found persistently and are considered central to the observed instability. For less-frequently associated nodes, there are several countries which show up repeatedly, but in different positions. These changes of macroscopic ordering are attributed to network effects, in the sense that the removal of some nodes can only lead to a rise in $\lambda_1^E$ when contained in certain compounds. It is most difficult, i.e.\ there are the fewest combinations found to lift $\lambda_1^E$,  in the beginning of 2008, suggesting that the instability is most deeply rooted in the network structure of the E-PIN at the onset of the GFC'08. In addition, any combination able of lifting  $\lambda_1^E$ at that time involves the group of nodes Guernsey, Egypt, Luxembourg and the UK, which is particularly interesting because Luxembourg is not frequently occurring, but highly central to the E-PIN. Note, furthermore, that the three countries mostly associated with the drop in $\lambda_1^E$ in 2005, Bahrain, Bermuda and Kuwait, are found to be related to a similar phenomenon within the SD-PIN (see \hyperref[si:9]{App.\ 3: S-9}), which underpins the obtained results and further demonstrates the applicability of the used concepts.\\
To quantify the concentration of countries classified as OFC among the nodes associated with the instability of the E-PIN, we estimate the average representation of OFC among \textit{all} nodes found by step two of the above search routine. We define the quotient 
\begin{equation}
Q_{OFC}^{\lambda}\left(n\right)\,=\,\frac{f^{\lambda}_{OFC}\left(n\right)}{f^{E}_{OFC}}\,,
\end{equation}
where $f^{x}_{OFC}\,,\,x\in \left\lbrace \lambda,\,E\right\rbrace$, is the averaged relative frequency of OFC between 2005-2008 in the groups associated with the $\lambda_1^E$-instability and in the E-PIN, respectively. A value of $Q_{OFC}^{\lambda}$ greater than \textit{one}, which is observed consistently, means a statistical over-representation of OFC among the countries associated with the $\lambda_1^E$-instability before the GFC'08. One observes a decrease of $Q_{OFC}^{\lambda}$, with rising $n$ which is expected because OFC represent the minority of nodes. \\
In our opinion, this consistent over-representation of OFC suggests the necessity of a further investigation of the central role of individual countries and groups of countries, such as OFC, regarding the stability of the global financial system, for which methodologies from network science, as presented in this work, could prove useful.\\
Two remarks are due at this point. First, even though we observed a structural destabilisation of the E-PIN, as measured by $\lambda_1^{E}$, as early as 2005, a note of caution should be made, when directly relating this finding to the GFC'08. It is not clear, if and how this instability played a role during the crisis, where the main question concerns the reconciliation of the static structural picture with a dynamic event. A low value of $\lambda_1^{E}$ might, however, be associated with a potentially increased susceptibility of the financial system towards shocks.\\[.1cm] \label{text:fed-rate}
Second, the algebraic connectivity of the LD-PIN, $\lambda_1^{LD}$, might as well serve as an indicator for potential financial crises because a separation of large parts of international debt markets is equally expected to lead to drastic consequences for the global financial system. In the current analysis, the temporal evolution of $\lambda_1^{LD}$ does not give reasons for major concerns. However, from the beginning of data taking, its value is constantly decreasing, still remaining at a relatively high level, till it suddenly rises in the beginning of 2006. A possible explanation for this behaviour, which will not be discussed further, is the rise of the Target Nominal Federal Funds Rate of the US Federal Reserve System \cite{fed_rep}. This coincides with a temporary slow-down in the increase of the number of edges in the LD-PIN (see \hyperref[fig:1]{Fig.\ 1-b}) and a halt in its expansion (see \hyperref[fig:3]{Fig.\ 3-a}).
\subsection{Indicator II: The Edge Density of the LD-PIN}
\label{sec:results-2.2}
\begin{center}
     \begin{figure}[!ht]
          \label{fig:4}
          \includegraphics[width=1.\textwidth]{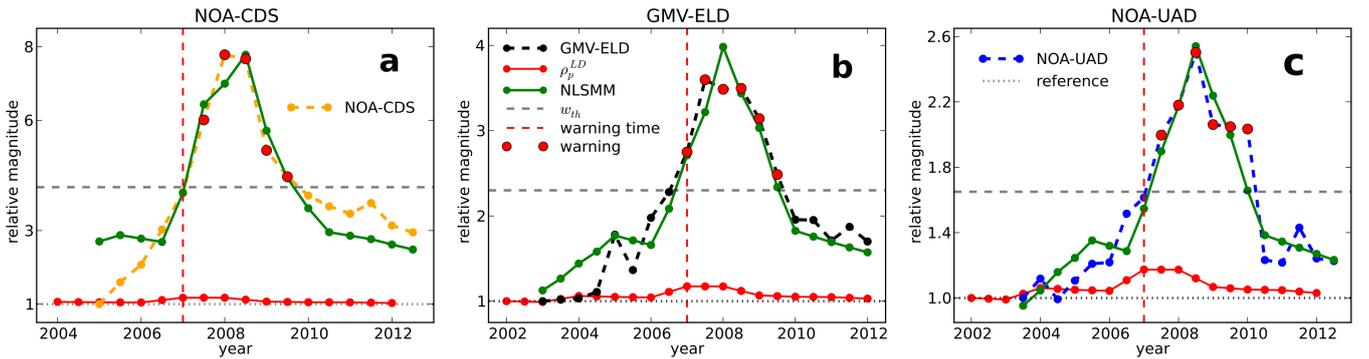}
          \caption{\small{Fit of the non-linear short-term memory model (NLSMM, Eq.\ \ref{eq:mm}) for the phenomenological description
          of the proliferation of OTC-traded financial derivative products, where values are given in relative magnitudes with respect 
          to a reference year. One mostly observes a good agreement between data and the NLSMM.
          Setting a warning threshold $w_{th}$, which has been set as a multiple $F_{max}^{D,GDP}$ of world-GDP (gray dashed line), 
          potentially dangerous levels of interdependency (red dots) can be detected, where the red dashed line indicates the time of first warning. 
          For positive time shifts $\Delta t$,
          such a warning signal is generated before the actual fitted values reach $w_{th}$, as is the case for NOA-CDS (a) and NOA-UAD (c).
          For all shown derivatives, the NLSMM in combination with the specified warning thresholds $w_{th}$, produces warning signals 
          at the beginning of 2007. It is particularly interesting that the NLSMM describes the ``hidden'' variable NOA-UAD 
          well \cite{bis_der,bis_tri,isda_3}.
          \textbf{a}: NOA-CDS, with $\gamma_1=11.0$, $\gamma_2=6.6$, $\Delta t=6\,$months, $p_r=0.92$, $w_{th}=0.56\,$world-GDP. 
          \textbf{b}: GMV-ELD, with $\gamma_1=7.3$, $\gamma_2=8.0$,  $\Delta t=0$, $p_r=0.95$, $w_{th}=0.014\,$world-GDP.
          \textbf{c}: NOA-UAD, with unallocated derivatives, $\gamma_1=5.9$, $\gamma_2=6.0$,  $\Delta t=6\,$months, $p_r=0.95$, 
          $w_{th}=0.75\,$world-GDP.}}
     \end{figure}
\end{center}

Our second network indicator for potential financial crises is the \textit{edge density} of the LD-PIN at the minimal percolation edge threshold $\rho_{p}^{LD}=52\,$million\ USD. As shown in \hyperref[fig:1]{Fig.\ 1-b}, its temporal evolution can be used to describe the magnitudes of NOS-CDS and GMV-ELD \cite{bis_der}. It can be seen that the evolution of all three quantities share some prominent features, while the curves for NOA-CDS and GMV-ELD lag somehow behind. The fact that the scale of the rise and fall of the values of derivatives is much larger than that of $\rho_p^{LD}$ can be explained by several factors. First, financial derivatives can be re-bundled to higher-order products, inflating the total value with respect to the underlying base assets, resulting in a global derivative market (notional amounts) which is at least one order in magnitude larger than the total volume in PIN \cite{bis_der} or world-GDP \cite{wb_data}. Furthermore, it is known that small changes in (edge) density can cause abrupt changes in global properties of a wide range of systems (phase transition), including many complex networks \cite{net_attack,non_lin_per,sm_phase}. \\
Consequently, there must be non-linear effects present. We introduce a simple phenomenological \textit{non-linear short-term memory model} (NLSMM), which is based on the assumption that total market values of financial derivatives scale non-linearly with the percolation edge density of the LD-PIN, $\rho_p^{LD}$, while certain memory (hysteresis) effects can be observed.\\ 
For the total market value $V_{D}\left(t_n\right)$ (NOA or GMV) of some derivative $V_{D}$ at time $t_n$, with respect to a reference value $V_r=V_D\left(t_r\right)$, we write
\begin{equation}
V_{D}\left(t_n\right)\,=\,V_{r}\cdot a_r\,\left[\,\bar{\rho}^{\gamma_1}\left(t_n\right)\,+\,\bar{\rho}^{\gamma_2}\left(t_{n-1}\right)\,\right],
\quad\mbox{where}\quad
\bar{\rho}\left(t_n\right)\,\equiv \,\rho_{p}^{LD}(t_n)\,/\,\rho_{p}^{LD}(t_r)\,.
\label{eq:mm}
\end{equation}
Here, $t_{n-1}$ denotes the data point before $t_n$, which is the previous year for a 1-year period of the CPIS data \cite{cpis};
$\gamma_1$ and $\gamma_2$ are scaling exponents, where a value different from one implies non-linearity; $a_r$ is a scaling factor, accounting for the arbitrarily chosen reference value $V_r$.\\
We consider different lead/lag time shifts of $\Delta t = +12,+6,0,-6,-12$\ months, according to the 6-months period of derivative data \cite{bis_der}, between $\rho_p^{LD}$ and NOA-CDS/GMV-ELD, when performing the fits to determine the scaling exponents $\gamma_{1,2}$ (see section \hyperref[sec:meth]{Methodology} for details on the fitting procedure.). Positive values of $\Delta t$ (lead) indicate that changes in $\rho_p^{LD}$ \textit{cause} changes in NOA-CDS/GMV-ELD, while negative values (lag) indicate that changes in $\rho_p^{LD}$ \textit{are caused} by changes in NOA-CDS/GMV-ELD. We find $\left(\Delta t\right)_{CDS} = 6\,$months and $\left(\Delta t\right)_{ELD} = 0\,$months. As expected, there is a tendency towards a lead of $\rho_p^{LD}$ with respect to NOA-CDS/GMV-ELD, while a higher resolution in both data sets \cite{cpis,bis_der} is expected to considerably clarify these relations. Note furthermore that positive values of $\Delta t$ enhance the applicability of the NLSMM (\ref{eq:mm}) for the indication of financial crises, since potentially high levels of market interdependence can be spotted earlier.\\
The results for both fits, using (\ref{eq:mm}), are shown in \hyperref[fig:4]{Figs.\ 4-a and\ -b} with reference values of 2005 and 2002, respectively. One can see that the NLSMM achieves a good description of the proliferation of CDS and ELD in most cases and that the agreement is  especially good for the years around the GFC'08 between 2007-2009. The mismatch for the year-2005 NOA-CDS point may either be considered as an outlier, or be explained through the coinciding start of the CDS statistics data taking \cite{bis_der}, which fell into a time of large market growth.\\
For both fits, the scaling exponents $\gamma_{1,2}$ are much larger than \textit{one}, showing that the relations between $\rho_p^{LD}$ and the market values of financial derivative products are highly non-linear. Note that changes in the reference point $V_r$ mainly change the coefficient $a_r$, while the scaling exponents $\gamma_{1,2}$ stay approximately constant. We define the ratio $m\equiv\gamma_{2}/\gamma_{1}$, which indicates the time span over which changes in the LD-PIN and derivatives markets are coupled, where a value greater than one means that past-year values of $\rho_p^{LD}$ contribute stronger in (\ref{eq:mm}) than present-year values. It might serve as a measure for ``market memory''. Here, $m_{ELD}\gg m_{CDS}$, indicating that changes in $\rho_p^{LD}$ have a longer lasting effect on GMV-ELD than on NOA-CDS, which means, on the other hand, that NOA-CDS reacts faster to changes in the LD-PIN.\\
A way to make use of the NLSMM for the quantitative indication of systemic risk and, as such, potential financial crises, is to set a warning threshold $w_{th}$ in terms of a maximal level for the market value of a certain product, which one deems still safe. One such possibility is a dynamic threshold, i.e.\ a threshold which allows for market changes which a certain derivative product may be coupled to, such as a macroeconomic reference variable (RV). 
If $V_{RV}\left(t\right)$ denotes the value of an RV at time $t$, we require
\begin{equation}
\label{eq:wth}
V_D\left(t\right)\,\leq\,w_{th}\left(t\right)\,=\,F^{D,RV}_{max}\cdot\,V_{RV}\left(t\right)\,,
\end{equation}
where $V_D$ is the market value of some derivative product (NOA or GMV), and $F^{D,RV}_{max}$ is a multiplier, which sets the maximal proportion between $V_D$ and $V_{RV}$. This is a dynamic threshold, in the sense that, if $V_{RV}$ increases, $V_{D}$ is also allowed to increase. However, if $V_{RV}$ decreases, such that $V_D$ comes close to $w_{th}$ or if  $V_D$ exceeds the threshold, $V_D$ must be reduced by some mechanism. A regulatory approach, based on monetary incentives to enforce a certain ``equilibrium level'' $F^{D,RV}_{max}$, could be to imposed an RV-progressive transaction or holding tax on financial derivatives, which makes especially speculative trading unattractive, as soon as the level of derivative $D$ comes close to $F^{D,RV}_{max}$.\\
In \hyperref[fig:3]{Fig.\ 3-e}, we investigate eight macroeconomic quantities with respect to their usability as RV for GMV-ELD, where the relative magnitude of $C_{ref}\equiv V_{ELD}/V_{RV}$ is plotted with reference to their year-2002 values (foreign exchange market turnover volumes for North America, FX vol. (N.A.), are only available from October 2004 on).\\
We indicate the shape which a suitable reference curves $C_{ref}$ should approximately follow (shaded area in \hyperref[fig:3]{Fig.\ 3-e}), to deliver a \textit{clear} warning signal before the GFC'08, i.e.\ to be able to set $w_{th}$ appropriately. We identify four potential RV, namely world-GDP \cite{wb_data}, global trade volume of goods (world trade vol., \cite{comtrade}), global stock of foreign direct investment (world-FDI stock, \cite{unctad}) and the foreign exchange market turnover volume in Singapore (FX vol. (SG), \cite{fx_vol_sg}), while we suggest to exclude the latter because it only covers a regional market. In our opinion, world-GDP offers the best RV for the case at hand. In addition, GDP is among the most-frequently referred-to indicators to quantify economic activity and, as such, represents a well-established baseline which the OTC-derivatives market may be compared to.\\
The value of $F^{D,RV}_{max}$ does not need to be set precisely, in the sense that equivalent warning signals will be generated within a comfortably large range of values. Taking world-GDP as RV, $F^{D,GDP}_{max}$ can be set within the ranges of $0.53-0.85$ and $0.012-0.018$ for warning signals in the beginning of 2007 for NOA-CDS and GMV-ELD, respectively, while the actual values have been set to $0.56$ and $0.014$ in \hyperref[fig:4]{Figs.\ 4-a and\ -b}. This might provide enough time to prevent the worst consequences of the insetting downturn.\\
To further investigate the relation between the LD-PIN and the OTC-derivatives market and the applicability of the proposed methodology, we tested the NLSMM (\ref{eq:mm}) for the description of all major classes of financial derivatives and their first subcategories, given in \cite{bis_der}. The results are summarised in \hyperref[tab:2]{Tab.\ 2}. One can see that the NLSMM gives mostly good descriptions for CDS and ELD, while those for foreign exchange (FXD) and interest rate  derivatives (IRD) are generally poor. Interestingly, NOA of unallocated derivatives (UAD, see \hyperref[fig:4]{Fig.\ 4-c}), which stems from the difference of reporting institutions between \cite{bis_der} and \cite{bis_tri,isda_3}, is described very well, offering a tool to indirectly measure this large, but ``hidden'' variable.\\
An additional class of derivatives where the NLSMM offers decent descriptive power and which saw a strong rise in popularity in the years before the GFC'08, for both hedging and speculation, are commodity-linked derivatives (CLD). Here, the strategy to hedge equity and bond risks using CLD, creating large amounts of additional interdependencies in global financial markets, failed in face of the crisis \cite{com_der}.\\[.1cm]
One might ask if the edge density of the E-PIN $\rho^{E}$ contributes additional information to the analysis similar to that provided by $\rho_p^{LD}$.
Note, at this point, that there is an underlying reason for choosing $\rho_p^{LD}$ to describe the proliferation of financial derivative products; namely, LD-securities can be seen as ``durable'' base assets for or be linked to various derivatives, which is generally not the case for E-securities. We may, however, infer some information from the evolution of $\rho^{E}$. We observe two major decreases of $\rho^{E}$ between the years from 2003 to 2004 and 2005 to 2006 (see \hyperref[fig:1]{Fig.\ 1-a}), where the number of edges is approximately increasing in the same manner as the number of nodes, consequently leading to a reduction of the edge density (see \hyperref[si:1]{App.\ 3: S-1}). These two events might be associated with the pick-up of equity investment after the 2001 recession \cite{dot_com} and the slow down of growth of the LD-PIN (see \hyperref[text:fed-rate]{previous subsection}), offering an incentive for investors to switch from debt to equity securities.\\[.2cm]
In summary, both early-warning indicators for financial crises are intrinsically tied to the network perspective, where the global financial system is treated as a complex multi-layered network consisting of strongly-interacting components. Especially, in view of the recent financial crisis, which conventional models have not been able to foresee, or even to describe properly, this offers a completely new perspective to macroeconomics, where the topology of a complex system, such as the international financial architecture, is taken into account explicitly. In our opinion, the GFC'08 is an excellent learning ground for the development of new methodologies aiming at the prevention of future large-scale economic downturns. We finally stress that, one of the biggest obstacles for the development of such tools is the scarcity of openly-available high-quality data, which is especially daunting in the dawning age of \textit{big data}.
\section{Discussion}
\label{sec:dis}
We have investigated the applicability of the presented early-warning indicators for financial crises by looking at the robustness of the results against changes in the edge threshold $e_{th}$, which has previously been set according to the percolation properties of the LD-PIN (see Section\ \hyperref[sec:meth]{Methodology}). It turns out that results from the E-PIN are highly insensitive to the choice of $e_{th}$, while results stemming from the LD-PIN are robust against variations of $e_{th}$ in a conveniently large window of up to $30\,$million\ USD.\\
To probe the threshold dependence of the results, we rise $e_{th}$ in a step-wise fashion from $1\,$million to $1\,$billion\ USD, considering a total of 500 values, which are equally-spaced on a logarithmic scale. The results from this exercise are shown in \hyperref[si:6]{App.\ 3: S-6}.\\
The threshold dependences of the network sizes, in terms of the numbers of nodes of the E- and LD-PIN, are shown in panels (a) and (b), respectively, where a near constant value of $e_{th}^p=52\,$million\ USD is identified above which the LD-PIN (b) disintegrates rapidly.\\
The $e_{th}$-dependence of the edge density $\rho^{LD}$ of the LD-PIN, which we have taken as a network-proxy to measure the interdependence of financial markets, is shown in panel (d) of \hyperref[fig:S4]{Fig.\ S-4} in \hyperref[si:7]{App.\ 3: S-7}. The detailed evolution of $\rho^{LD}$ over time is rather sensitive to the choice of $e_{th}$, while most fitting results for the description of financial derivatives, when using (\ref{eq:mm}), do not change within a range of $e_{th}=25-55\,$million\ USD (shaded regions in panels (b) and (d) of \hyperref[fig:S4]{Fig.\ S-4}), where the number of nodes of the LD-PIN is approximately constant. This window of possible values is considered comfortably large for practical implementations of the proposed methodologies. There are, however, slight differences of the goodness-of-fit for different derivatives. The proliferation of CDS is generally described better at larger threshold within the stated region, while one achieves better description of ELD and CLD for lower values of $e_{th}$.\\
The threshold dependence of the algebraic connectivity $\lambda_1^E$ of the E-PIN, as a measure for robustness of the global financial system, and as such for the world economy, is shown in panel (c) of \hyperref[fig:S4]{Fig.\ S-4}. The main features of the temporal evolution of $\lambda_1^E$, such as its sharp drop in 2005 and its persistent low value until the GFC'08, are largely \textit{threshold-independent}. No qualitative changes are observed for values of $e_{th}$ between $1$ to $100\,$million\ USD, which makes the algebraic connectivity of the E-PIN a robust indicator for practical implementation in the current case. We remark that the actual details of such a $\lambda_1^E$-instability should be carefully investigated, when detected, as we have done in this work. Its large value for high thresholds $e_{th}$ stems from the induced removal of nodes, resulting in a more robust core-component.\\
Considering the anti-correlation of volume fractions of the total PIN shown in \hyperref[fig:2]{Fig.\ 2-b}, one may also take an alternative approach: a joint threshold of, say, $e_{th}=110\,$million\ USD, for both PIN together, where particular values used for the E- and LD-PIN are determined on a yearly basis according to their respective fractions of the total PIN. This has been done. It turns out that results do not change considerably, which is the reason for choosing a simpler single value for both networks at all times.
\section{Methodology}
\label{sec:meth}
\textbf{Network set-up:} Data for different types of cross-border portfolio investment networks (PIN) for eleven consecutive years, 2002-2012 (beginning of the year), come from the Coordinated Portfolio Investment Survey (CPIS) \cite{cpis}, conducted by the International Monetary Fund (IMF). Portfolio investment is an indirect investment, defined here as a cross-border transactions or positions of equity or debt securities, excluding direct investment, reserve assets, financial derivatives and bank loans \cite{bpm6}. The CPIS includes aggregated data for end-of-the-year positions in USD from 78 reporting creditor countries for three types of securities: equity (E), long- (LD) and short-term (SD) debts. E-securities, such as shares, stocks, participations or similar documents indicating ownership, and LD-securities, such as bonds, debentures and notes with a maturity of more than one year, make up for the majority in terms of total volume, approximately $94\,\%$, as seen in \hyperref[fig:3]{Fig.\ 3-a} and \hyperref[si:3]{App.\ 3: S-3}. We therefore concentrate on these two types. A summary of results on the SD-PIN and how the proposed methodologies can be used for its analysis is given in \hyperref[si:9]{App.\ 3: S-9}.\\ 
In the networks constructed from the CPIS data, nodes are represented by individual countries and weighted directed edges by aggregated portfolio investment positions between them. In terms of the weight matrix $W$ (see \hyperref[si:1]{App.\ 3: S-1}), an investment position from country $i$ in country $j$ is given by $w_{ij}$.\\
To allow for a better comparability of results over time, all monetary values (investment positions, derivatives statistics, trade flows and market turnover volumes) have been adjusted for changes in world-GDP, using the GDP deflator \cite{wb_data} (constant 2013-values).\\
The LD-PIN is seen to have an approximately constant \textit{percolation edge threshold} with an minimum value of $e_{th}^{p}=52\,$million\ USD (see \hyperref[fig:S4]{Fig.\ S-4-b} in \hyperref[si:8]{App.\ 3: S-8}). This is the edge weight above which the LD-PIN rapidly disintegrates, when consecutively removing edges of larger weights, as measured by the size of its largest strongly-connected component. There are two notable exceptions to this behaviour. Namely, before and during the GFC'08, in 2007 and 2008, the percolation point rises by about a factor of three, which is an interesting observation in its own right. The graph connectivity at the edge threshold $e_{th}^{p}$, where edges $w_{ij}<e_{th}^{p}$ are deleted, i.e.\ $w_{ij}\equiv 0$, is expected to contribute dominantly to the global properties of the weighted network, while still allowing for a good comparability of results over time.\\
Data between non-reporting countries, as well as liabilities between any two countries, are missing from the CPIS \cite{cpis}, which gives the resulting networks a ``sea urchin-like'' topology. Nevertheless, the present data include mutual investment positions for all major economies, bar the Mainland of China, where cross-border capital flows are highly restricted and, as such, would not contribute significantly \cite{imf_china_rep,ch_safe_stats}.\\
To account for both the percolation properties of the LD-PIN and the incompleteness of data, we define a PIN as the largest strongly-connected component, after applying the minimum percolation edge threshold $e_{th}^{p}=52\,$million\ USD.\\
There is no such general percolation point for the E-PIN. However, for some years, a rather low value below $10\,$million\ USD can be found, while for other instances, a more continuous disintegration with a rising edge threshold is observed, as shown in \hyperref[fig:S4]{Fig.\ S-4-a}. This is most notable for the year 2007, where the largest number of nodes for low values of $e_{th}$ is observed, which already indicates a relatively fragile expansion of the E-PIN prior to the GFC'08. We note that it may come to negative positions in the CPIS data under certain conditions, as explained in the CPIS guide \cite{cpis}. Such positions are indeed observed, making a total number of nodes larger than the $78$ reporting countries possible.\\ 
For reasons of comparability and simplicity, the same edge threshold is used for the E-PIN in this investigation, where the network volume is of similar magnitude. Final PIN volumes, after extracting the largest strongly connected components, are around $95\,\%$ of the initial amounts, i.e.\ we make use of the great majority of data after the above-described clean-ups.\\[.2cm]
\textbf{Model fit:} We use a least-square method to fit the percolation edge density of the LD-PIN $\rho^{LD}_p$ to the data from the OTC-derivatives statistics \cite{bis_der}, when implementing the non-linear short-term memory model (NLSMM,\ Eq. \ref{eq:mm}). The best-fit lead/lag time shifts between both quantities, where we considered values of $\Delta t\in\lbrace+12,+6,0,-6,-12\rbrace\,$months, are obtained by minimising the normalised squared-difference $||V_D-fit(V_D,\Delta t)||^2/||V_D||^2$, where $||\cdot||$ is the Euclidean norm of a vector containing values from different time instances, $V_D$ is the market value of financial derivative $D$, in either notional outstanding amounts (NOA) or gross-market values (GMV), and $fit(V_D,\Delta t)$ is the best-fit of (\ref{eq:mm}) for a given $\Delta t$. Overall differences in the average values of derivatives for different time windows are accounted for in this way.\\
The Pearson product correlation coefficient $p_r$ \cite{prob_book1} has been used as a goodness-of-fit criterion, where we accept the best-fit $fit(V_D,\Delta t)$ if $p_r(V_D,fit(V_D,\Delta t))\geq 0.9$, which turns out to provide overally-good results. We conditionally accept a fit if $0.85\leq p_r(V_D,fit(V_D,\Delta t))< 0.9$, while we reject a fit if $p_r(V_D,fit(V_D,\Delta t))< 0.85\,$.\\
Note that PIN and derivatives data have different time resolutions of 12 and 6 months, respectively. To be able to make full use of all available data, we interpolate $\rho^{LD}_p$ to obtain the same resolution for both data sets. This is justified on the ground that $\rho^{LD}_p$ is a slowly-changing variable compared to all derivatives data. Alternatively, dropping one half of the derivatives data is seen to lead to less accurate results, in the sense that more fits are actually accepted due to the induced smoothing-effects.\\
As can be seen from \hyperref[tab:2]{Tab.\ 2}, one obtains one negative scaling exponent (mostly $\gamma_1$) for the best-fit solution in some cases. Depending on the absolute value, as compared to the other exponent, this means that the corresponding contribution is suppressed, and the fit is effectively 
achieved through one parameter only. Note, however, that fit results are not satisfactory in all of these cases.\\
Data for credit default swaps (CDS) are only available starting from the end of 2004 onward. To be able to make use of all data points for all times, we consider intervals of varying lengths for different values of $\Delta t$, which introduces a bias towards lags of $\rho^{LD}_p$, when selecting the best-fit. Obtaining persistent  \textit{lead} relations for all given CDS statistics, strengthens the results, which does not change when considering intervals of equal lengths.
\section*{Acknowledgements}
This research was supported by the Hong Kong Research Grants Council under the GRF Grant CityU1109/12. The authors also thank the developers and supporters of the Python programming language, including \cite{ipython} and \cite{nx}, which have been used throughout this study.

%
\addcontentsline{toc}{section}{References}
{\small{

}}
\clearpage
\section*{Appendix\ 1: Figures}
\addcontentsline{toc}{section}{Appendix\ 1: Figures}
\label{app:fig}
\begin{center}
     \begin{figure}[!ht]
          \label{fig:1}
          \includegraphics[width=1.\textwidth]{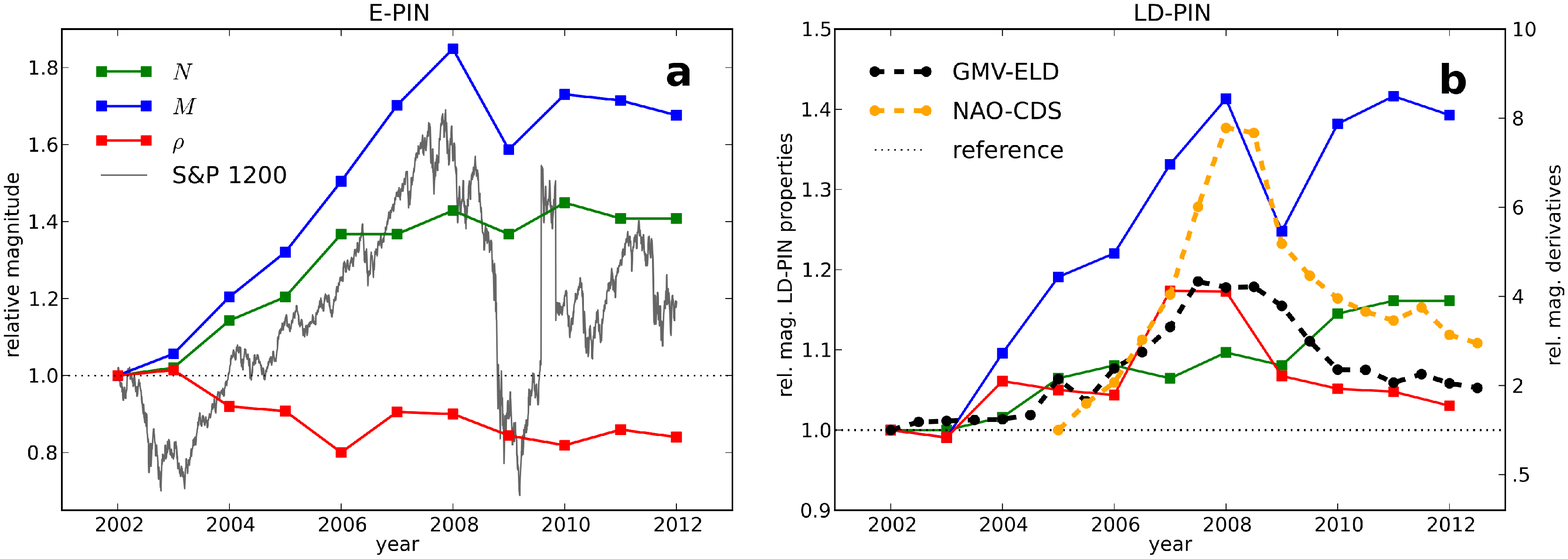}
          \caption{\small{Temporal Evolution of network key properties: number of nodes $N$, 
          number of edges $M$ and the resulting edge density $\rho$, 
          of the two considered PIN relative to their corresponding year-2002 values (dotted reference line).
          \textbf{a}: E-PIN. The daily time series of the $S\&P\;\mathrm{Global}\,1200$ 
          stock index is given for reference.
          \textbf{b}: LD-PIN. The temporal evolution of the gross-market value of equity-linked derivatives 
          (GMV-ELD) and the total notional outstanding amount of credit default swaps (NOA-CDS, data not 
          available prior to 2005) mirror the qualitative evolution of the percolation edge density $\rho_p^{LD}$.}}
     \end{figure}
\end{center}
\begin{center}
     \begin{figure}[!ht]
          \label{fig:2}
          \includegraphics[width=1.\textwidth]{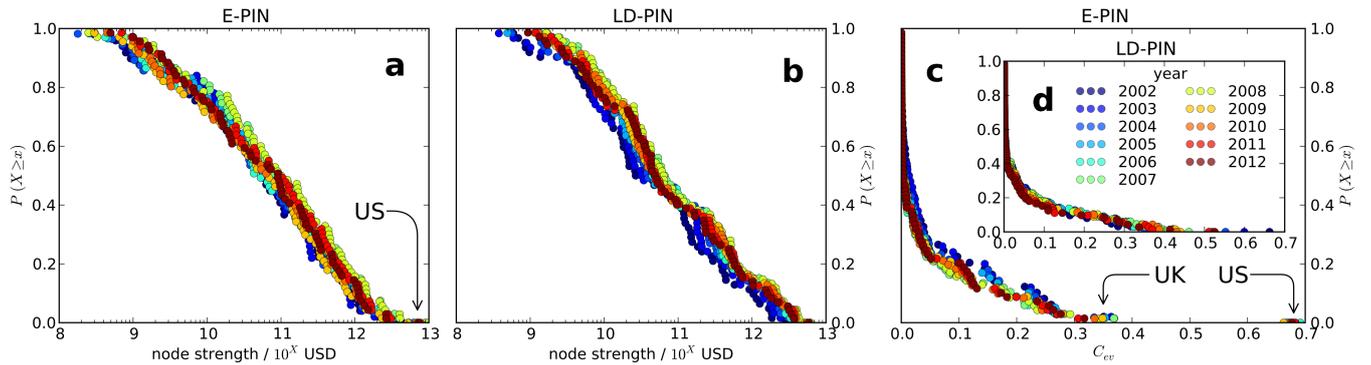}
          \caption{\small{Cumulative node strength (\textbf{a}, \textbf{b}) and eigenvector centrality (\textbf{c}, \textbf{d}) 
          distributions of the E- (\textbf{a}, \textbf{c}) and the LD-PIN (\textbf{b}, \textbf{d}),
          respectively. Numerical values cover about the same ranges for both networks. In particular, node strengths span about five 
          orders of magnitude with the largest values before the GFC'08. Strength distributions may be classified as ``super-heavy-tailed'', since 
          there is a $\mathcal{O}\left(1\right)$-probability for most values which indicates a strong hierarchical network structure.
          The US is by far the most central node in the E-PIN, with the highest strength and a dominating eigenvector centrality.}}
     \end{figure}
\end{center}
\begin{center}
     \begin{figure}[!ht]
          \label{fig:3}
          \includegraphics[width=1.\textwidth]{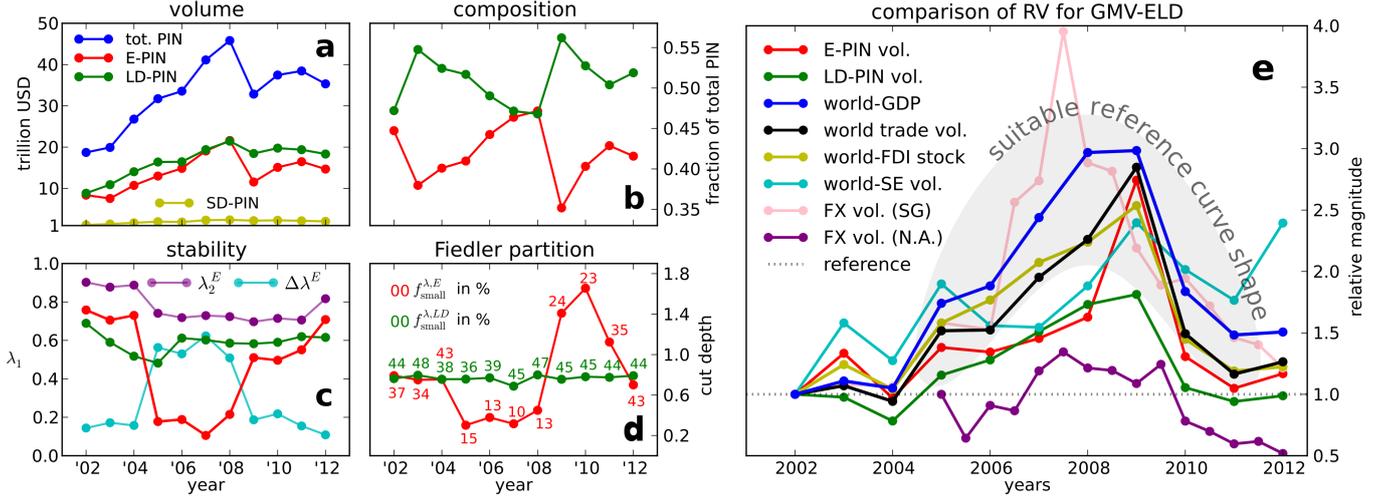}
          \caption{\small{Left: Temporal evolution of general PIN characteristics. 
          \textbf{a}: Volumes (edge weight sums) in USD of the total PIN, the E- and the LD-PIN. 
          For better comparability, all monetary values have been adjusted for changes in world-GDP 
          (constant year-2013 values). The GFC'08 causes a contraction of both networks, which is especially pronounced in the E-PIN.
          \textbf{b}: Fractions of total PIN volume from panel (a) contained in the E- and LD-PIN. One observes a constant shift in composition
          from the LD-PIN towards the E-PIN in the years before crisis.
          \textbf{c}: Algebraic connectivity $\lambda_1$ as a measure for network robustness against node/edge failure (removal). 
          $\lambda_1^E$ reaches an all-time low just before the GFC'08. $\lambda_2^E\geq\lambda_1^E$ and 
          $\Delta\lambda^E=\lambda_2^E-\lambda_1^E$ are given as an alternative reference scale to evaluate the stability of the E-PIN.
          \textbf{d}: Cut depths of Fieder bi-sections of the E- and the LD-PIN. Numbers represent the fractions $f_{\mathrm{small}}^{\lambda}$ 
          in percent of nodes contained in the 
          corresponding smaller sections. While the LD-PIN is cut evenly without large variations 
          in cut depth, the corresponding quantities of the E-PIN point to major topological changes before and during the crisis.
          Right (\textbf{e}): Comparison of potential macroeconomic reference variables (RV) for setting a dynamic warning threshold $w_{th}(t)$ for 
          the gross-market value (GMV) of OTC-traded equity-linked derivatives (ELD): E- and LD-PIN volumes, world-GDP \cite{wb_data}, 
          world trade volume (goods) \cite{comtrade}, global stock of foreign direct investment (FDI) \cite{unctad}, global stock exchange 
          trading volume \cite{wb_data}, foreign exchange market turnover volumes in Singapore and North America (US, Canada and Mexico, only available
          from October 2004 on). For a RV $X$, the temporal evolution of the quantity GMV-ELD/$X$ is given relative to its year-2002 value,
          where nearest points in time have been matched. The shaded area indicates the shape a suitable RV should follow to 
          generate a clear warning signal prior to the GFC'08. World-GDP (blue line) is seen to offer the best RV from the given sample.}}
     \end{figure}
\end{center} 
\clearpage
\section*{Appendix\ 2: Tables}
\addcontentsline{toc}{section}{Appendix\ 2: Tables}
\label{app:tab}
\begin{table}[hb!]
\renewcommand{\arraystretch}{1.3}
\begin{center}
\begin{tabular}{|c||c|c| L{13cm} |}
\hline
n \cellcolor[gray]{0.9} & $p_{\lambda}$ \cellcolor[gray]{0.9} &$Q_{OFC}^{\lambda}$ \cellcolor[gray]{0.9} 
 & countries most-frequently associated with the $\lambda_1^E$-instability \cellcolor[gray]{0.9}\\
\specialrule{.1em}{.05em}{.05em} 
$1$ & - & - & Bahrain, Kuwait \\
\hline
$2$ & - & - & Bermuda, Guernsey \\
\hline
$3$ & - & - & Bermuda, Guernsey, Jersey, Egypt \\
\specialrule{.1em}{.05em}{.05em} 
$4$ & $0.03$ & $2.05$  & Bermuda, Guernsey, United Kingdom, Jersey, Egypt, Ukraine, Aruba, Luxembourg, Turkey, Latvia, Isle of Man, United States, Spain, Malaysia, Cura\c{c}ao \\
\hline
$5$ & $0.09$ & $1.78$  & Bermuda, Guernsey, United Kingdom, Jersey, Egypt, Russia, Uruguay, Switzerland, Turkey, Slovenia, Slovak Republic, Romania, Isle of Man, Hong Kong, Cura\c{c}ao \\
\hline
$6$ & $0.11$ & $1.61$  & Bermuda, Guernsey, United Kingdom, Jersey, Malta, Russia, Egypt, Mexico, Kazakhstan, Venezuela, Philippines, Ukraine, Czech Republic, Italy, Israel \\
\hline
$7$ & $0.23$ & $1.52$  & Bermuda, Guernsey, United Kingdom, Jersey, Egypt, Denmark, Isle of Man, Luxembourg, India, Mexico, Netherlands, Singapore, Slovak Republic, Cyprus, Venezuela \\
\hline
$8$ & $0.24$ & $1.39$  & Bermuda, Guernsey, United Kingdom, Jersey, Egypt, Latvia, Canada, Austria, Bulgaria, Spain, Malta, Germany, Belgium, Switzerland, Gibraltar  \\
\hline
$9$ & $0.40$ & $1.33$  & Bermuda, Guernsey, United Kingdom, Jersey, Egypt, India, Gibraltar, Latvia, Spain, Germany, Luxembourg, Kazakhstan, Philippines, Italy, Japan \\
\hline
$10$ & $0.41$  & $1.36$  & Bermuda, Guernsey, United Kingdom, Jersey, Egypt, India, Latvia, Hungary, Singapore, Canada, Barbados, Sweden, Netherlands, Spain, Bahamas\\
\hline
\end{tabular}
\caption{\small{List of countries most-associated with the $\lambda_1^E$-instability for groups of sizes $n=1$-$10$. A country or group of countries is associated with the $\lambda_1^E$-instability if its removal from the E-PIN leads to a partial or complete lifting of $\lambda_1^E$ over a value of $0.5$ during the period 2005-2008. The cases $n=1,2,3$ have been treated exhaustively. Bahrain and Kuwait are seen to be the only single countries whose removal lifts $\lambda_1^E$, while the picture turns out to be more complex when considering larger groups. For $n=4$-$10$, a two-step breadth-first search algorithm has been used, excluding Bahrain and Kuwait. Results have been averaged over five full rounds. The average probability of finding a combination of nodes, whose removal leads to a jump in  $\lambda_1^E$, is given by $p_{\lambda}$. Its step-like behaviour for rising values of $n$ is caused by the rounding rule used in step two of the search routine. The 15 most-frequently found countries are shown. The group consisting of Bermuda, Guernsey, United Kingdom, Jersey and Egypt is  persistently associated with the $\lambda_1^E$-instability, suggesting that these countries are most central for its formation. $Q_{OFC}^{\lambda}$ estimates the average representation of off-shore financial centres (OFC) among all nodes found by the search routine. A value larger than one, which is observed persistently, means a relative over-representation of OFC.}}
\end{center}
\label{tab:1}
\end{table}
\clearpage
\begin{table}[t!]
\begin{center}
\small{
\renewcommand{\arraystretch}{1.1}
\begin{tabular}{|l|l|l||r|r|r|r|r|r|r|l|}
\hline
\multicolumn{3}{|l||}{ \cellcolor[gray]{0.9} name} & \multicolumn{1}{c|}{\cellcolor[gray]{0.9} $f_{GDP}^{\mathrm{world}}$} 
 & \multicolumn{1}{c|}{\cellcolor[gray]{0.9} $a_r$ } & \multicolumn{1}{c|}{\cellcolor[gray]{0.9} $\gamma_1$} 
 & \multicolumn{1}{c|}{\cellcolor[gray]{0.9} $\gamma_2$ } & \multicolumn{1}{c|}{\cellcolor[gray]{0.9} $m$} 
 & \multicolumn{1}{c|}{\cellcolor[gray]{0.9} $\Delta\,t$}  & \multicolumn{1}{c|}{\cellcolor[gray]{0.9} $p_r$} 
 & \multicolumn{1}{c|}{\cellcolor[gray]{0.9}  decision} \\
\specialrule{.1em}{.05em}{.05em}
\multirow{21}{.7cm}{NOA} &
\multicolumn{2}{l||}{\textit{total}} & 10.939 & 0.7 & 4.1 & 3.3 & 0.8 & 12 & 0.7 & no\\
\cline{2-11}
 & \multirow{3}{.7cm}{CDS} & \textit{total} & 0.934 & 0.9 & 11.0 & 6.6 & 0.6 & 6 & 0.92 & \textbf{yes}\\
 &  & SNI & 0.543 & 0.7 & 9.6 & 6.9 & 0.7 & 6 & 0.93 & \textbf{yes}\\
 &  & MNI & 0.39 & 1.6 & 12.8 & 6.0 & 0.5 & 6 & 0.9 & \textbf{yes}\\
\cline{2-11}
 & \multirow{3}{.7cm}{FXD} & \textit{total} & 1.024 & 0.6 & 3.4 & 2.6 & 0.8 & 6 & 0.64 & no\\
 &  & F$\&$S & 0.52 & 0.6 & -1.3 & 5.7 & - & -6 & 0.8 & no\\
 &  & SWP & 0.265 & 0.8 & 0.1 & 2.6 & 17.8 & 12 & 0.31 & no\\
 &  & OPT & 0.239 & 0.6 & 5.7 & 2.1 & 0.4 & 6 & 0.89 & maybe\\
\cline{2-11}
 & \multirow{3}{.7cm}{IRD} & \textit{total} & 7.454 & 0.8 & 2.7 & 3.8 & 1.4 & 12 & 0.62 & no\\
 &  & FWD & 0.64 & 1.1 & -4.1 & 6.3 & - & 12 & 0.53 & no\\
 &  & SWP & 5.803 & 0.8 & 2.8 & 3.7 & 1.3 & 12 & 0.62 & no\\
 &  & OPT & 1.011 & 0.7 & 5.6 & 3.2 & 0.6 & 6 & 0.85 & maybe\\
\cline{2-11}
 & \multirow{3}{.7cm}{ELD} & \textit{total} & 0.166 & 0.7 & 4.6 & 7.0 & 1.5 & -6 & 0.92 & \textbf{yes}\\
 &  & F$\&$S & 0.043 & 0.8 & 8.0 & 2.4 & 0.3 & 6 & 0.84 & no\\
 &  & OPT & 0.122 & 0.6 & 4.7 & 6.2 & 1.3 & -6 & 0.9 & \textbf{yes}\\
\cline{2-11}
 & \multirow{3}{.7cm}{CLD} & \textit{total} & 0.215 & 0.7 & 9.8 & 14.1 & 1.4 & -6 & 0.87 & maybe\\
 &  & GLD & 0.011 & 0.4 & 5.5 & -0.7 & - & 0 & 0.77 & no\\
 &  & OTH & 0.205 & 1.0 & 10.6 & 15.0 & 1.4 & -6 & 0.87 & maybe\\
 &  & F$\&$S  & 0.123 & 0.6 & 15.4 & 8.0 & 0.5 & 6 & 0.89 & maybe\\
 &  & OPT & 0.082 & 0.9 & 15.0 & 13.4 & 0.9 & -6 & 0.85 & maybe\\
\cline{2-11}
 & \multicolumn{2}{l||}{UAD} & 1.146 & 0.5 & 5.9 & 6.0 & 1.0 & 6 & 0.95 & \textbf{yes}\\
\specialrule{.1em}{.05em}{.05em}
\multirow{19}{.7cm}{GMV} &
 \multicolumn{2}{l||}{\textit{total} } & 0.331 & 0.6 & 3.3 & 7.2 & 2.2 & 12 & 0.77 & no\\
\cline{2-11}
 & \multirow{3}{.7cm}{CDS} & \textit{total} & 0.052 & 0.9 & 17.8 & 16.4 & 0.9 & 12 & 0.94 & \textbf{yes}\\
 &  & SNI & 0.031 & 0.6 & 17.5 & 17.2 & 1.0 & 12 & 0.94 & \textbf{yes}\\
 &  & MNI & 0.021 & 2.1 & 18.2 & 14.9 & 0.8 & 12 & 0.93 & \textbf{yes}\\
\cline{2-11}
 & \multirow{3}{.7cm}{FXD} & \textit{total} & 0.037 & 0.4 & 4.3 & 5.7 & 1.3 & 12 & 0.72 & no\\
 &  & F$\&$S & 0.013 & 0.5 & -9.9 & 6.4 & - & 0 & 0.66 & no\\
 &  & SWP & 0.017 & 0.5 & 2.3 & 5.2 & 2.2 & 12 & 0.67 & no\\
 &  & OPT & 0.006 & 0.5 & 7.0 & 7.4 & 1.1 & 12 & 0.88 & maybe\\
\cline{2-11}
 & \multirow{3}{.7cm}{IRD} & \textit{total} & 0.151 & 0.8 & -4.4 & 7.1 & - & 12 & 0.62 & no\\
 &  & FWD & 0.001 & 0.7 & 5.5 & 10.6 & 1.9 & 12 & 0.66 & no\\
 &  & SWP & 0.131 & 0.8 & -5.1 & 7.2 & - & 12 & 0.62 & no\\
 &  & OPT & 0.018 & 0.8 & 0.3 & 6.2 & 19.3 & 12 & 0.62 & no\\
\cline{2-11}
 & \multirow{3}{.7cm}{ELD} & \textit{total} & 0.019 & 0.6 & 7.3 & 8.0 & 1.1 & 0 & 0.95 & \textbf{yes}\\
 &  & F$\&$S & 0.005 & 0.5 & 8.2 & 7.8 & 0.9 & 6 & 0.93 & \textbf{yes}\\
 &  & OPT & 0.014 & 0.6 & 8.0 & 7.5 & 0.9 & 0 & 0.95 & \textbf{yes}\\
\cline{2-11}
 & \multirow{3}{.7cm}{CLD} & \textit{total}  & 0.036 & 0.9 & 10.0 & 15.8 & 1.6 & 0 & 0.86 & maybe\\
 &  & GLD & 0.001 & 0.5 & 5.2 & 4.4 & 0.8 & -6 & 0.68 & no\\
 &  & OTH & 0.035 & 1.2 & 10.6 & 16.7 & 1.6 & 0 & 0.87 & maybe\\
\cline{2-11}
 & \multicolumn{2}{l||}{UAD} & 0.037 & 0.5 & -2.6 & 7.2 & - & 6 & 0.81 & no\\
\cline{2-11}
 & \multicolumn{2}{l||}{GCE} & 0.063 & 0.4 & 4.5 & 4.1 & 0.9 & 12 & 0.85 & maybe\\
\specialrule{.1em}{.05em}{.05em}
\multicolumn{2}{|l|}{NOA-EXD} & FUT & 0.423 & 0.6 & 3.5 & 5.6 & 1.6 & -12 & 0.9 & \textbf{yes}\\
\multicolumn{2}{|l|}{\textcolor{white}{NOA-EXD}} &  OPT & 0.809 & 0.8 & 5.5 & 4.2 & 0.8 & -6 & 0.8 & no\\
\hline
\end{tabular}
\caption{\small{Fit results between the percolation edge density $\rho_p^{LD}$ of the LD-PIN and notional outstanding amounts (NOA) and gross-market values (GMV) of all major classes of OTC-traded financial derivatives and their first subcategories \cite{bis_der}, using the NLSMM (\ref{eq:mm}) (scaling factor $a_r$ and exponents $\gamma_{1,2}$ for a fit of $V_D(t)/V_r$). The fraction/multiple of world-GDP $f_{GDP}^{\mathrm{world}}$ of NOA/GMV as of the middle of 2008 is given for reference for each class/subcategory, where the mean between the 2008- and 2009-values has been taken for world-GDP. The quantity $m\equiv \gamma_2/\gamma_1$ indicates the time span over which changes in the LD-PIN and the derivative market are coupled (``market memory''), where a value greater than one means that past-year values of $\rho_p^{LD}$ contribute stronger in (\ref{eq:mm}) than present-year values. $\Delta t$ (in months) is the best-fit lead/lag time shift between $\rho_p^{LD}$ and NOA/GMV of a certain derivative class, where positive values indicate a lead of $\rho_p^{LD}$. We use the Pearson product correlation coefficient $p_r$ between the best-fit and market values of derivative products as a goodness-of-fit criterion, where we accept a fit if $p_r\geq 0.9\,$. We say that a certain product may be described by the NLSMM if $0.9>p_r\geq 0.85$ (conditional acceptance), and reject the fit if $p_r<0.85\,$. The major classes are credit default swaps (CDS), foreign exchange derivatives (FXD), interest rate derivatives (IRD), equity-linked derivatives (ELD) and commodity-linked derivatives (CLD). The first sub-categories are single- and multi- name instruments SNI and MNI for CDS, respectively, and forwards and swaps (F$\&$S), swaps (SWP), options (OPT) and forwards (FWD) for the other classes. CLD additionally include gold derivatives (GLD) and others commodities (OTH). Unallocated derivatives (UAD) are values which are not covered in \cite{bis_der}, but are included in \cite{bis_tri,isda_3}. Gross-credit exposure (GCE) measures the positive net-value of contracts, after mutual obligations have been set off (netting). Fit results for NOA of exchange-traded derivatives (EXD) are given for reference, where one sees that future contracts (FUT) can also be described by the NLSMM. The NLSMM \ref{eq:mm} is seen to be especially suitable for the description of the proliferation of CDS (NOA and GMV), which is the financial derivative product which has most-frequently been related to the GFC'08.}}
}
\end{center}
\label{tab:2}
\end{table}
\clearpage
\section*{Appendix\ 3: Supplementary Information}
\addcontentsline{toc}{section}{Appendix\ 1: Supplementary Information}
\label{app:si}
\subsection*{S-1: Relevant Concepts and Notations from Graph Theory}
\addcontentsline{toc}{subsection}{S-1: Relevant Concepts and Notations from Graph Theory}
\label{si:1}
Graph theory provides a general mathematical framework to represent and quantify complex networks and their properties \cite{NW_book1,NW_book2,NW_book3}. We will use the words network and graph synonymously. A weighted and directed network can be represented by a graph $G\,=\,\left(\mathcal{V},\,\mathcal{E}\right)$, where $\mathcal{V}\,=\,\lbrace v_1,\, \dots,\, v_N \rbrace$ is the set of $N\geq 2$ nodes in the graph, and $\mathcal{E}\left(w_{ij}>0|\,i,j\in\lbrace 1,\,\dots,\,N\rbrace\right)$ is the set of weighted edges from node $v_i$ to node $v_j$, with $M\,=\,ord(\mathcal{E})$ denoting the number of edges irrespective of their weights. The whole graph can be represented by a real weight matrix, $W\,=\,[w_{ij}]\,\in\,\mathbb{R}^{N\times N}$ ($w_{ij}\,\neq\,w_{ji}$, in general). We do not allow self-loops, i.e.\ $w_{ii}\,=\,0$ for all $i\,\in\,\lbrace 1,\,\dots,\,N\rbrace$. We define the \textit{volume} $V$ of a weighted network as the sum of all its edge weights, which is $V^G=\sum_{i}\sum_{j}\,w_{ij}$,  $\lbrace i,j\rbrace\,\in\,\lbrace 1,\,\dots,\,N\rbrace$.\\
The in- and out-\textit{degrees} of a node $i$ are the numbers of in-coming and out-going connections, respectively. In the weight matrix $W$, these are given by the respective number of non-zero entries in the $i^{th}$ column or row. The out- and in-\textit{strengths} of that node are defined as the $i^{th}$ row- and column-sums of $W$, respectively. The total strength/degree of a node is then the sum of its in- and out-degrees/strengths. \\
The \textit{edge density} of a graph is defined as the number of actual edges $M$ divided by the number of maximally possible edges $M_{max}$. For directed networks, 
$$\rho\,=\,\frac{M}{M_{max}}\,=\,\frac{M}{N^2-N}\,.$$ 
It is a network size-independent measure for the average connectivity between nodes, but does not take the network structure into account. \\
One says that node $i$ connects to node $j$ if there exists a directed path from $i$ to $j$. A graph is said to be  \textit{strongly-connected}, if such a directed path exists between any pair of nodes $\left(i,\,j\right)$. The weight matrix $W$ is \textit{irreducible} in this case. If such a path does not exit for all node pairs, but the underlying undirected graph is connected, the network is said to be \textit{weakly-connected}.\\
The importance of a node to a network is often evaluated in terms of the so-called \textit{centrality} measures, which quantify the extent to which a node participates in the path structure of a network, taken from a certain point of view \cite{network_flow,graph_theoretic_centrality,comp_cent}. Simple examples are node degree and strength which measure short-range connectivity, but often scale with other centrality measures \cite{c_corr,rn_wtw_1,rn_wtw_2}. A measure which captures a node's higher-order and recursive connections within the whole network is \textit{eigenvalue centrality}. It is a type of feedback centrality, defined as the node-component of the eigenvector corresponding to the largest eigenvalue of the weight matrix $W$.\\
The \textit{algebraic connectivity} $\lambda_1$ of a graph is here defined as the real-part of the smallest non-zero eigenvalue of the normalised Laplacian $L_N\,=\,\mathbbm{1}-\,D_{in}^{-1}\cdot W^T$ \cite{lam_0}, where $D_{in}^{-1}\,=\,diag(1/s_{1}^{in},\,\dots,\,1/s_{N}^{in})$ is the diagonal matrix consisting of all nodes' in-strengths. The Laplacian matrix has exactly one zero eigenvalue for every strongly-connected component. For the case of a single strongly-connected component, $\lambda_1$ is the second smallest eigenvalue of $L_N$. It is a measure for the \textit{robustness} of a network against edge/node removal \cite{NW_book1,lam_2,lam_4} because a zero-value means the decomposition of the network into two disconnected components. It can be used to derive a lower bound on the minimum \textit{cut ratio} $R_{cut}(S) = w_{cut}(s_1,s_2)/(|s_1|\cdot|s_2|)$ for the bi-section $S(s_1,s_2)$ of a graph into two disconnected partitions of nodes $s_1$ and $s_2$ with $|s_1|$ and $|s_2|$ nodes, respectively, where $w_{cut}$ is the weight sum of removed edges to perform the cut \cite{NW_book1,lam_5,lam_3}. Finding the optimal bi-section $S(s_1,s_2)$ which minimises $R_{cut}$ is an NP-hard problem. However, the \textit{Fiedler vector} $\vec{v}_1$, which is the eigenvector of $L_N$ corresponding to $\lambda_1$, can be used to approximate this optimal bi-section, if $\lambda_1$ is small \cite{com_dec_1,GT_book2}. The sections $s_+$ and $s_-$ are then given by the nodes corresponding to the positive and negative entries of $\vec{v}_1$, respectively (Fiedler bi-section), where $f_{\mathrm{small}}^{\lambda}\equiv min\lbrace |s_+|,|s_-|\rbrace/N$ denotes the fraction of nodes contained in the smaller section. Note that a Fiedler bi-section, like any bi-section where the section sizes are not fixed, is \textit{not} expected to be \textit{balanced} ($|s_+|\approx |s_-|$ or $f_{\mathrm{small}}^{\lambda}\approx0.5$).\\ 
A low value of $\lambda_1$, sufficiently close to zero, is assumed to indicate a structural instability. If the corresponding Fiedler bi-section is balanced, this instability/fragility of the network is of \textit{global} nature, i.e.\ affecting the majority of nodes. On the contrary, an unbalanced bi-section points to a \textit{partial} instability, such as a sparsely connected group of nodes.\\
There is no absolute level for $\lambda_1$ to evaluate the structural stability of a given network. For practical applications, e.g.\ for the analysis of PIN, where we have a single strongly-connected component and $N\gg 2$, the inequality \cite{lam_0}, 
$$0\,<\,\lambda_1\,\leq\,\frac{N}{N-1}\,\approx\,1$$
offers a \textit{quasi-absolute scale} to judge the potential fragility of a network or compare results for different instances over time with varying numbers of nodes and edges. A second possibility to assess the value of $\lambda_1$, which might prove even more useful for practical applications, is to compare it to $\lambda_2$, with $\lambda_1\leq\lambda_2$, which is the second smallest non-zero eigenvalue of $L_N$, taken by its real-part. If $\lambda_2$ is now roughly constant for different configurations of a network and considerably smaller than one, it can be used as an additional scale for the assessment of the values of $\lambda_1$ and their changes over time, where one can consider the quantity $\Delta\lambda\equiv\lambda_2-\lambda_1$.\\
In the context of weighted networks, we can define the \textit{cut depth} of a graph bi-section $S$, as $D_{cut}(S)\equiv f^S_w/f^S_M$, where $f^S_w$ and $f^S_M$ are the fractions of the network volume and of the number of edges connecting $s_1$ and $s_1$, respectively, i.e.\ the fractions of volume and number of edges which have to be removed to perform the cut. It is a parameter independent of both the network size and edge weight, which evaluates the average weight of the cut edges, when bi-sectioning a graph. Let $\braket{w(S)}$ and $\braket{w(G)}$ denote the average edge weight contained in $w_{cut}$ and in $G$, respectively. Then, the cut depth $D_{cut}(S)$ relates them, as $\braket{w(S)}=D_{cut}(S)\cdot \braket{w(G)}$. Consequently, when $D_{cut}(S)>1$, stronger-than-average edges need to be cut to perform the bi-section $S$, while the opposite is true if $D_{cut}(S)<1$. In \hyperref[si:6]{App.\ 3: S-6}, we show how the cut depths can be used to describe Fiedler graph bi-sections.\\
We introduce the parameter triple $T(G)=(\lambda_1,\,f_{\mathrm{small}}^{\lambda},\,D_{cut}^{\lambda})$ for the classification of Fiedler bi-sections of weighted single-component networks $G$. For Large values of $f_{\mathrm{small}}^{\lambda}>0.2-0.3$, one may say that an associated $\lambda_1$-instability is global, i.e.\ affecting the majority of nodes in a network, while small values point to a local instability, where the break-up of a network, e.g.\ through node failure, is expected to only affect a minority of nodes. Values of $D_{cut}^{\lambda,E}>1$, which indicate the disruption of large links in the event of a break-up, might have profound impacts on the dynamics within a network.\\
\subsection*{S-2: Country Lists}
\addcontentsline{toc}{subsection}{S-2: Country Lists}
\label{si:2}
Lists of 78 reporting countries of the Coordinated Portfolio Investment Survey (CPIS, \cite{cpis}) from the International Monetary Fund (IMF) as of end-2011, the IMF-46 off-shore financial centres (OFC) given in \cite{def_ofc} and the 24 OFC that can be found in the equity securities portfolio investment network (E-PIN) at least once.
\begin{table}[ht!]
\renewcommand{\arraystretch}{1.2}
\begin{center}
\begin{tabular}{|L{3cm} || L{13cm} |}
\hline
\cellcolor[gray]{0.9}group name & \cellcolor[gray]{0.9}included countries\\
\specialrule{.1em}{.05em}{.05em} 
CPIS Reporters & Argentina, Aruba, Australia, Austria, Bahamas, Bahrain, Barbados, Belgium, Bermuda, Brazil, Bulgaria, Canada, Cayman Islands, Chile, Hong Kong, Macao, Colombia, Costa Rica, Cura\c{c}ao, Cyprus, Czech Republic, Denmark, Egypt, Estonia, Finland, France, Germany, Gibraltar, Greece, Guernsey, Hungary, Iceland, India, Indonesia, Ireland, Isle of Man, Israel, Italy, Japan, Jersey, Kazakhstan, South Korea, Kosovo, Kuwait, Latvia, Lebanon, Lithuania, Luxembourg, Malaysia, Malta, Mauritius, Mexico, Netherlands, Netherlands Antilles, New Zealand, Norway, Pakistan, Panama, Philippines, Poland, Portugal, Romania, Russia, Singapore, Slovak Republic, Slovenia, South Africa, Spain, Sweden, Switzerland, Thailand, Turkey, Ukraine, United Kingdom, United States, Uruguay, Vanuatu, Venezuela \\
\hline
OFC (IMF-46) & Andorra, Anguilla, Antigua and Barbuda, Aruba, Bahamas, Bahrain, Barbados, Belize, Bermuda, Cayman Islands, Cook Islands, Costa Rica, Cyprus, Dominica, Gibraltar, Grenada, Guernsey, Hong Kong, Ireland, Isle of Man, Jersey, Lebanon, Liechtenstein, Luxembourg, Macao, Malaysia, Malta, Marshall Islands, Mauritius, Monaco, Montserrat, Nauru, Netherlands Antilles, Niue, Palau, Panama, Samoa, Seychelles, Singapore, St. Kitts and Nevis, St. Lucia, St. Vincent and the Grenadines, Switzerland, Turks and Caicos Islands, Vanuatu, Virgin Islands (UK)\\
\hline
OFC (E-PIN) & Aruba,  Bahamas,  Bahrain,  Barbados,  Bermuda,  Cayman Islands, Costa Rica,  Cyprus,  Gibraltar,  Guernsey,  Hong Kong,  Ireland,  Isle of Man,  Jersey,  Lebanon,  Luxembourg,  Macao,  Malaysia,  Malta,  Mauritius,  Netherlands Antilles,  Panama,  Singapore,  Switzerland\\
\hline
\multicolumn{2}{c}{ }\\
\multicolumn{2}{c}{Table\ S-2: Country lists.}
\end{tabular}
\end{center}
\label{tab:S1}
\end{table}
\clearpage
\subsection*{S-3: PIN Statistics}
\addcontentsline{toc}{subsection}{S-3: PIN Statistics}
\label{si:3}
PIN statistics corresponding to \hyperref[fig:1]{Figs.\ 1} and \hyperref[fig:3]{3-(a-d)}, and \hyperref[fig:S5]{Figs.\ S-5-(a-f)} of this SI are given. From top to bottom: Network volumes $V$ of the the total, E- and LD-PIN in trillion USD (Note that monetary values have been adjusted to year-2013 values, using the GDP-deflator \cite{wb_data}. Different PIN components do not sum up to the total PIN volume because the extraction of the largest strongly connected component after application of the edge threshold.), volume fractions $f_V$ as of the total PIN, number of nodes $N$, number of edges $M$, edge density $\rho$, algebraic connectivity $\lambda_1$ and $\lambda_2\geq\lambda_1$, fractions of nodes $f_{\mathrm{small}}^{\lambda}$ contained in the smaller partition of the Fiedler graph bi-section and the corresponding cut depths $D_{cut}^{\lambda}$ of Fiedler graph bi-sections.
\begin{table}[ht!]
\begin{center}
\renewcommand{\arraystretch}{1.4}
\begin{tabular}{|l||c|c|c|c|c|c|c|c|c|c|c|}
\hline
\cellcolor[gray]{0.9}year & \cellcolor[gray]{0.9}2002 & \cellcolor[gray]{0.9}2003 & \cellcolor[gray]{0.9}2004 & 
\cellcolor[gray]{0.9}2005 & \cellcolor[gray]{0.9}2006 & \cellcolor[gray]{0.9}2007 & \cellcolor[gray]{0.9}2008 & 
\cellcolor[gray]{0.9}2009 & \cellcolor[gray]{0.9}2010 & \cellcolor[gray]{0.9}2011 & \cellcolor[gray]{0.9}2012 \\
\specialrule{.1em}{.05em}{.05em} 
$V^{TOT}$ &  18.7  &  19.9  &  26.8  &  31.7  &  33.6  &  41.1  &  45.8  &  32.8  &  37.5  &  38.4  &  35.3 \\
$V^E$   &  8.4  &  7.6  &  10.7  &  13.0  &  14.9  &  19.1  &  21.6  &  11.6  &  15.1  &  16.5  &  14.7 \\
$V^{LD}$  &  8.8  &  10.9  &  14.0  &  16.4  &  16.5  &  19.4  &  21.4  &  18.5  &  19.8  &  19.4  &  18.3 \\
$V^{SD}$   &  1.2  &  1.4  &  1.7  &  2.0  &  1.9  &  2.3  &  2.4  &  2.2  &  2.3  &  2.2  &  2.0 \\
\hline
$f_V^{E}$  &  0.447  &  0.380  &  0.401  &  0.410  &  0.443  &  0.464  &  0.472  &  0.352  &  0.403  &  0.429  &  0.416 \\
$f_V^{LD}$  &  0.472  &  0.548  &  0.524  &  0.517  &  0.490  &  0.472  &  0.468  &  0.562  &  0.527  &  0.504  &  0.519 \\
$f_V^{SD}$  &  0.062  &  0.069  &  0.062  &  0.062  &  0.056  &  0.056  &  0.052  &  0.067  &  0.061  &  0.056  &  0.057 \\
\hline
$N^E$  &  49  &  50  &  56  &  59  &  67  &  67  &  70  &  67  &  71  &  69  &  69 \\
$N^{LD}$  &  62  &  62  &  63  &  66  &  67  &  66  &  68  &  67  &  71  &  72  &  72 \\
$N^{SD}$  &  54  &  55  &  56  &  60  &  59  &  61  &  64  &  61  &  67  &  61  &  63 \\
\hline
$M^E$ &  886  &  936  &  1067  &  1170  &  1333  &  1508  &  1638  &  1406  &  1533  &  1519  &  1485 \\
$M^{LD}$ &  1189  &  1178  &  1303  &  1416  &  1451  &  1583  &  1680  &  1484  &  1643  &  1684  &  1656 \\
$M^{SD}$  &  622  &  688  &  742  &  785  &  816  &  873  &  890  &  827  &  850  &  848  &  827 \\
\hline
$\rho^E$   &  0.38  &  0.38  &  0.35  &  0.34  &  0.30  &  0.34  &  0.34  &  0.32  &  0.31  &  0.32  &  0.32 \\
$\rho^{LD}$  &  0.31  &  0.31  &  0.33  &  0.33  &  0.33  &  0.37  &  0.37  &  0.34  &  0.33  &  0.33  &  0.32 \\
$\rho^{SD}$  &  0.22  &  0.23  &  0.24  &  0.22  &  0.24  &  0.24  &  0.22  &  0.23  &  0.19  &  0.23  &  0.21 \\
\hline
$\lambda_1^E$  &  0.76  &  0.71  &  0.73  &  0.18  &  0.19  &  0.11  &  0.21  &  0.51  &  0.50  &  0.55  &  0.71 \\
$\lambda_2^E$  &  0.90  &  0.88  &  0.89  &  0.74  &  0.72  &  0.73  &  0.72  &  0.70  &  0.71  &  0.71  &  0.82 \\
$\lambda_1^{LD}$  &  0.69  &  0.59  &  0.52  &  0.48  &  0.61  &  0.60  &  0.59  &  0.58  &  0.59  &  0.62  &  0.62 \\
$\lambda_2^{LD}$  &  0.88  &  0.66  &  0.66  &  0.69  &  0.78  &  0.64  &  0.73  &  0.63  &  0.62  &  0.79  &  0.80 \\
$\lambda_1^{SD}$  &  0.42  &  0.62  &  0.65  &  0.56  &  0.60  &  0.53  &  0.65  &  0.48  &  0.13  &  0.44  &  0.62 \\
$\lambda_2^{SD}$  &  0.67  &  0.62  &  0.65  &  0.67  &  0.69  &  0.61  &  0.75  &  0.48  &  0.55  &  0.48  &  0.65 \\
\hline
$f_{\mathrm{small}}^{\lambda,E}$ &  0.37  &  0.34  &  0.43  &  0.15  &  0.13  &  0.10  &  0.13  &  0.24  &  0.23  &  0.35  &  0.43 \\
$f_{\mathrm{small}}^{\lambda,LD}$ &  0.44  &  0.48  &  0.38  &  0.36  &  0.39  &  0.45  &  0.47  &  0.45  &  0.45  &  0.44  &  0.44 \\
$f_{\mathrm{small}}^{\lambda,SD}$  &  0.22  &  0.27  &  0.34  &  0.42  &  0.31  &  0.48  &  0.48  &  0.48  &  0.04  &  0.41  &  0.40 \\
\hline
$D_{cut}^{\lambda,E}$  &  0.79  &  0.75  &  0.75  &  0.30  &  0.38  &  0.32  &  0.45  &  1.41  &  1.66  &  1.12  &  0.70 \\
$D_{cut}^{\lambda,LD}$  &  0.76  &  0.80  &  0.76  &  0.76  &  0.77  &  0.69  &  0.80  &  0.76  &  0.78  &  0.78  &  0.79 \\
$D_{cut}^{\lambda,SD}$  &  1.59  &  1.29  &  1.05  &  0.72  &  1.08  &  0.93  &  1.03  &  0.52  &  0.54  &  0.43  &  0.81 \\
\hline
\multicolumn{12}{c}{ }\\
\multicolumn{12}{c}{Table\ S-2: PIN statistics.}
\end{tabular}
\end{center}
\label{tab:S2}
\end{table}
\clearpage
\subsection*{S-4: Cumulative Edge Weight Distributions}
\addcontentsline{toc}{subsection}{S-4: Cumulative Edge Weight Distributions}
\label{si:4}
The edge weight distributions of the E- and LD-PIN are both strongly right-skewed and can be classified as heavy-tailed. Global characteristics, like the total range of covered values, are similar for both networks, spanning four orders of magnitude in edge weights and three orders in probability. There are, however, differences in the shapes of distributions. The edge weight distribution of the E-PIN drops faster than the one of the LD-PIN for small weights, while the opposite is true for large weights. In accordance with the results from node strength and eigenvector centrality, this emphasises the more pronounced hierarchical structure of the E-PIN, as compared to the LD-PIN. Moreover, overall edge weights in the E-PIN are found to be largest prior to the 2008 global financial crisis (GFC'08), which is in agreement with the strong increase of the network's volume at that time.
\begin{center}
     \begin{figure}[!ht]
          \label{fig:S1}
          \includegraphics[width=.49\textwidth]{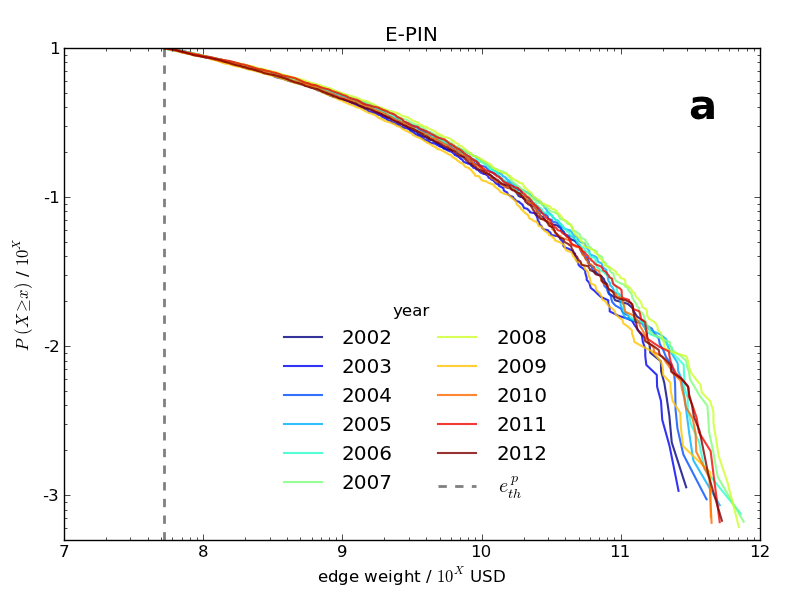}
          \includegraphics[width=.49\textwidth]{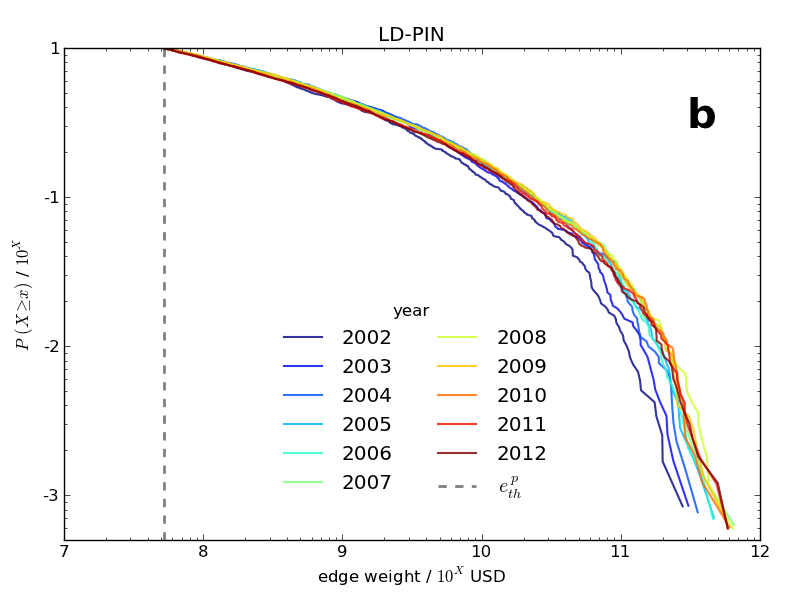}
     \end{figure}
     Figure\ S-1: Cumulative edge weight distributions. \textbf{a:} E-PIN. \textbf{b:} LD-PIN. 
\end{center}
\subsection*{S-5: Cross-border Portfolio Investment and International Trade}
\addcontentsline{toc}{subsection}{S-5: Cross-border Portfolio Investment and International Trade}
\label{si:5}
International trade volume is a main macroeconomic indicator for the state of the world economy. We compare the total volumes of international trade flows \cite{comtrade} between CPIS \cite{cpis} reporting countries and portfolio investment positions (a) and effective flows (b) for the E-PIN and the LD-PIN, where trade data are available for all CPIS reporters, except Guernsey, Isle of Man, Jersey and Kosovo. Effective flows here means the difference between consecutive investment positions between two countries, but does not necessarily imply a cash flow. This is especially thought to be true for the 2008 peak in equity flows (b), which is associated with a global drop in stock markets during the GFC'08 (see \hyperref[fig:1]{Fig.\ 1-a}), rather than a large amount of cross-border transactions. Portfolio investment positions and trade flows are of the same order in magnitude, while investment positions are, on average, a factor $1.4$ larger than trade flows on the given time span. By comparing (a) and (b), one can see that start-of-year investment positions are much better indicators than investment flows for the total volume of trade flows. When considering the impact of the GFC'08 on the trade flow volume, its drastic reduction of about $30\,\%$ is represented by a contraction of the E-PIN by $47\,\%$ and of the LD-PIN by $14\,\%$. Noting that the re-bounce of trade after the crisis is better matched by the E-PIN, we conclude that its volume offers a better proxy than the LD-PIN for international trade.
\begin{center}
     \begin{figure}[!ht]
          \label{fig:S2}
          \includegraphics[width=.49\textwidth]{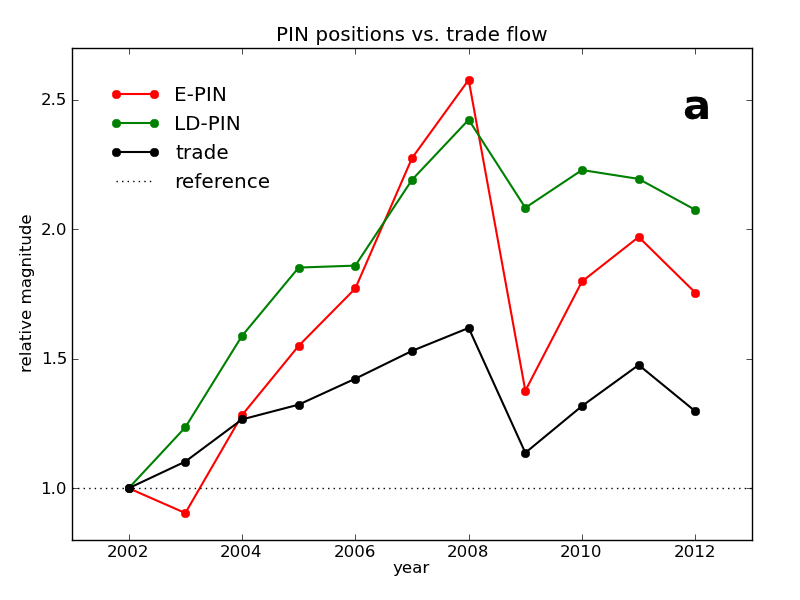}
          \includegraphics[width=.49\textwidth]{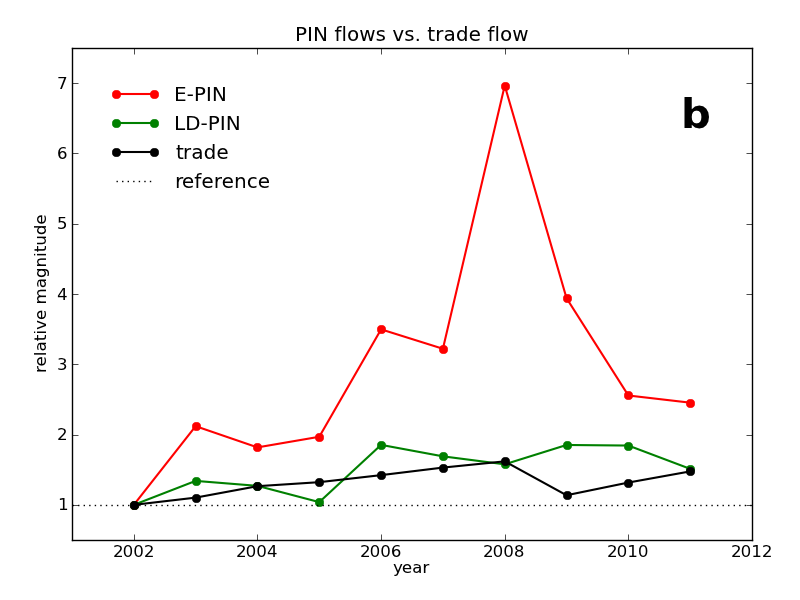}
     \end{figure}
     Figure S-2: Cross-border portfolio investment and international trade. \textbf{a:} Investment positions. \textbf{b:} Investment flows.
\end{center}
\subsection*{S-6: Cut Depths of PIN Random Bi-sections}
\addcontentsline{toc}{subsection}{S-6: Cut Depths of PIN Random Bi-sections}
\label{si:6}
We compare the average cut depths of different types of random bi-sections of the E- and the LD-PIN. For a random bi-section, the set of nodes of either network at a specific time is divided randomly into two disjoint groups and the cut depth $D_{cut}$ for edges connecting these two groups is calculated. Three different drawing methods are used for partitioning the sets of nodes, where one group is chosen accordingly and the other group is given by the remaining nodes: Fixed partition size containing half of the nodes, random size between $1$ and $N-1$ nodes, and same partition sizes as the corresponding Fiedler bi-section (see \hyperref[si:6]{App.\ S-6} and \hyperref[tab:S2]{Tab.\ S-2}). \\
For every year, network and drawing method, $10^4$ hypothetical bi-sections are performed and averages are taken. The results are summarised in \hyperref[fig:S3]{Fig.\ S-3}, where $99\,\%$-confidence intervals (CI) have been included. The cut depths of the corresponding Fiedler bi-sections are given for reference. When cutting a network in half (balanced bi-section), the expected cut depth is \textit{one} because contributions from heavy and light edges, independent of the edge weight distribution, balance each other, which is exactly what we observe (red lines). However, when partition sizes are random, the average partition will be unbalanced and one is, on average, more likely to cut light edges in the face of a heavily right-tailed edge weight distribution, as is the case for both PIN (see \hyperref[fig:S1]{Fig.\ S-1}). Consequently, one expects $D_{cut}<1$, which is again what we observe (blue lines). The characteristics of both Fiedler-like random bi-sections (green lines) can now be explained by combining the properties of balanced and unbalanced random partitions. For the E-PIN at the beginning and the end of the observation period, and for the LD-PIN during the whole observation period, Fiedler bi-sections are rather balanced (see \hyperref[tab:S2]{Tab.\ S-2}), leading to approximately unit cut depths of the corresponding random partitions. The Fiedler bi-section of the E-PIN is, however, unbalanced for the middle of the observation period, especially during the years 2005-2008, when the average cut depth of random bi-sections is dropping accordingly.\\
We see now that the Fiedler bi-sections (black lines), which minimise the cut ratio approximatively, are \textit{significantly} different from all random partitions, indicating their ability to identify efficient bi-sections. This, in turn, may be used to describe structural instabilities as indicated by low values of the algebraic connectivity $\lambda_1$. As described in the main text, the triple $T(G)=(\lambda_1,\,f_{\mathrm{small}}^{\lambda},\,D_{cut}^{\lambda})$ can be used to classify the Fiedler bi-sections and, if perceived as such, also the corresponding structural instability.
\begin{center}
     \begin{figure}[!ht]
          \label{fig:S3}
          \includegraphics[width=.49\textwidth]{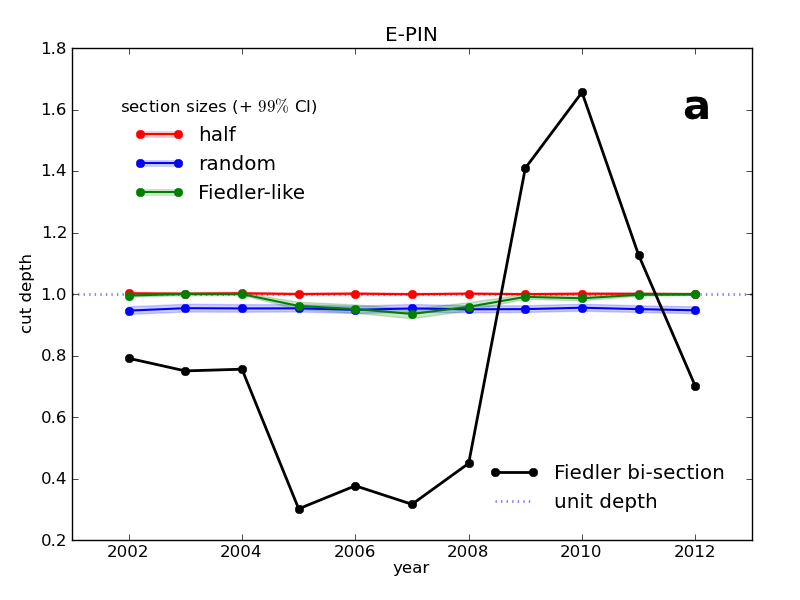}
          \includegraphics[width=.49\textwidth]{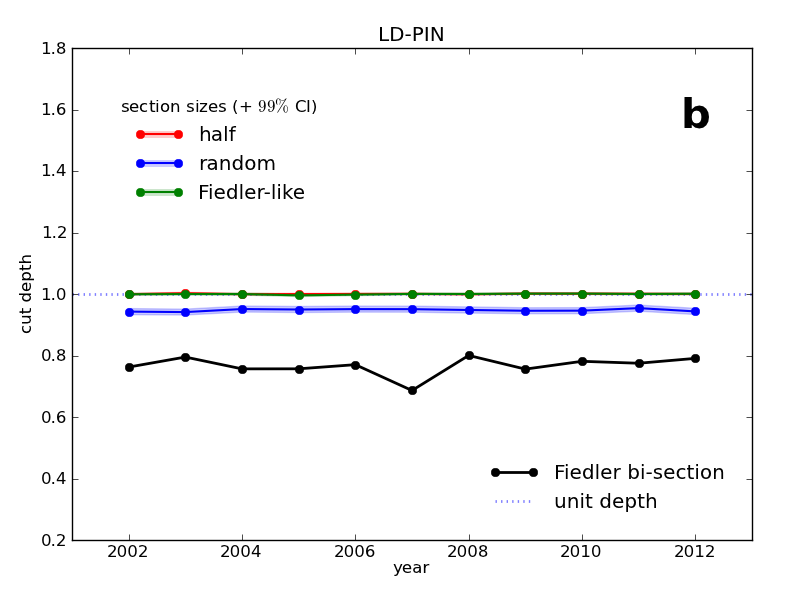}
     \end{figure}
     Figure S-3: Cut depths of PIN random bi-sections. \textbf{a:} E-PIN. \textbf{b:} LD-PIN.
\end{center}
\clearpage
\subsection*{S-7: E-PIN Sections $s_{\mathrm{small}}^{\lambda,E}$}
\addcontentsline{toc}{subsection}{S-7: E-PIN Sections $s_{\mathrm{small}}^{\lambda,E}$}
\label{si:7}
Parameter triple $T(E)=(\lambda_1^E,\,f_{\mathrm{small}}^{\lambda,E},\,D_{cut}^{\lambda,E})$ for the classification of Fiedler bi-sections and lists of countries contained in the smaller sections $s_{\mathrm{small}}^{\lambda,E}$ for all years. The numbers in parentheses are the in- and out-degrees of each node, respectively.\\
\begin{table}[ht!]
\begin{center}
\renewcommand{\arraystretch}{1.4}
\begin{tabular}{|c||c|c|c|| L{12cm} |}
\hline
\renewcommand{\arraystretch}{1.2}
\cellcolor[gray]{0.9} year & \cellcolor[gray]{0.9}$\lambda_1^E$ & \cellcolor[gray]{0.9}$f_{\mathrm{small}}^{\lambda,E}$ & 
\cellcolor[gray]{0.9}$D_{cut}^{\lambda,E}$ & \cellcolor[gray]{0.9}countries in $s_{\mathrm{small}}^{\lambda,E}$ \\
\specialrule{.1em}{.05em}{.05em} 
2002 & 0.76 & 0.37 & 0.79 & Argentina (9, 3), Austria (20, 29), Belgium (21, 29), Cyprus (4, 4), Czech Republic (8, 7), Egypt (4, 2), Finland (24, 17), France (31, 30), Germany (30, 31), Hungary (13, 2), Iceland (3, 4), Ireland (29, 29), Italy (27, 36), Luxembourg (34, 42), Netherlands (33, 33), Poland (15, 1), Spain (28, 22), Switzerland (27, 35) \\
\hline
2003 & 0.71 & 0.34 & 0.75 & Austria (18, 28), Belgium (24, 31), Chile (7, 4), Czech Republic (10, 7), Egypt (4, 1), Finland (25, 19), France (33, 34), Germany (31, 32), Hungary (16, 2), Ireland (31, 33), Italy (29, 40), Luxembourg (36, 41), Macao (1, 6), Netherlands (32, 36), Portugal (19, 18), Spain (27, 21), Switzerland (29, 36) \\
\hline
2004 & 0.73 & 0.43 & 0.75 & Argentina (7, 7), Austria (21, 34), Barbados (1, 3), Belgium (23, 38), Brazil (26, 13), Czech Republic (16, 9), Denmark (21, 37), France (34, 41), Germany (35, 39), Greece (22, 11), Hungary (16, 2), Iceland (3, 8), Ireland (33, 38), Italy (27, 41), Jersey (15, 36), Lebanon (1, 4), Luxembourg (41, 47), Macao (1, 8), Netherlands (34, 42), Poland (17, 1), Portugal (21, 20), Slovak Republic (2, 2), Spain (26, 23), Uruguay (1, 1) \\
\hline
2005 & 0.18 & 0.15 & 0.30 & Bahrain (3, 8), Cayman Islands (34, 7), Cyprus (2, 5), Egypt (9, 2), Guernsey (18, 43), Kuwait (2, 11), Lebanon (3, 4), Malaysia (21, 5), Turkey (20, 1) \\
\hline
2006 & 0.19 & 0.13 & 0.38 & Bahamas (18, 4), Bahrain (2, 11), Cayman Islands (38, 5), Egypt (12, 1), Kuwait (2, 13), Lebanon (4, 4), Malaysia (24, 6), Malta (3, 5), Turkey (21, 1) \\
\hline
2007 & 0.11 & 0.10 & 0.32 & Bahamas (23, 3), Bahrain (4, 16), Egypt (15, 3), Kuwait (2, 16), Lebanon (4, 6), Malta (7, 5), Sweden (33, 48) \\
\hline
2008 & 0.21 & 0.13 & 0.45 & Bahamas (23, 3), Bahrain (4, 10), Cayman Islands (39, 8), Egypt (18, 4), Kuwait (4, 17), Lebanon (2, 5), Malaysia (27, 18), Malta (12, 6), Turkey (26, 1) \\
\hline
2009 & 0.51 & 0.24 & 1.41 & Australia (33, 39), Bahamas (18, 3), Bahrain (5, 10), Bermuda (33, 32), Cayman Islands (42, 1), Egypt (17, 5), Isle of Man (10, 15), Kuwait (6, 16), Lebanon (5, 6), Malaysia (26, 15), Malta (8, 4), New Zealand (12, 7), Norway (22, 45), Poland (25, 11), United Kingdom (51, 54), United States (60, 57) \\
\hline
2010 & 0.50 & 0.23 & 1.66 & Bahamas (18, 4), Bahrain (5, 8), Cayman Islands (37, 3), Egypt (17, 5), Gibraltar (3, 4), Guernsey (26, 52), Iceland (4, 14), Isle of Man (10, 17), Jersey (32, 45), Kuwait (6, 16), Lebanon (5, 8), Malaysia (26, 16), Malta (10, 5), Turkey (24, 2), United Kingdom (54, 58), United States (64, 61) \\
\hline
2011 & 0.55 & 0.35 & 1.12 & Australia (34, 35), Bahamas (18, 6), Bahrain (5, 6), Bulgaria (3, 4), Canada (37, 50), Cayman Islands (40, 2), Egypt (17, 3), Gibraltar (3, 3), Guernsey (26, 53), Indonesia (27, 1), Isle of Man (15, 18), Israel (21, 16), Jersey (30, 45), Kuwait (4, 17), Lebanon (6, 6), Malaysia (26, 16), Malta (10, 5), New Zealand (15, 8), Panama (13, 1), Singapore (28, 36), Slovenia (4, 14), Turkey (25, 3), United Kingdom (49, 50), United States (64, 58) \\
\hline
2012 & 0.71 & 0.43 & 0.70 & Argentina (5, 4), Australia (32, 37), Barbados (1, 10), Bermuda (36, 27), Brazil (31, 18), Bulgaria (3, 4), Canada (38, 47), Cayman Islands (43, 3), Chile (16, 12), Colombia (14, 7), Denmark (25, 46), Greece (18, 14), Guernsey (27, 51), Hong Kong (33, 27), Iceland (3, 14), Indonesia (25, 1), Isle of Man (14, 20), Israel (19, 16), Japan (31, 46), Malaysia (26, 15), Mexico (21, 4), New Zealand (15, 13), Panama (14, 1), Singapore (27, 32), South Africa (24, 22), South Korea (24, 37), Switzerland (41, 48), United Kingdom (50, 56), United States (65, 58), Uruguay (1, 2) \\
\hline
\multicolumn{5}{c}{ }\\
\multicolumn{5}{c}{Tab.\ S-3: E-PIN section $s_{\mathrm{small}}^{\lambda,E}$.}
\end{tabular}
\end{center}
\label{tab:S3}
\end{table}
\clearpage
\subsection*{S-8: Threshold Dependence of Financial Crises Indicators}
\addcontentsline{toc}{subsection}{S-8: Threshold Dependence of Financial Crises Indicators}
\label{si:8}
Edge threshold ($e_{th}$) dependence of PIN network size by the number of nodes $N$ (a: LD-PIN, b: E-PIN)
and the two indicators for financial crises (c: $\lambda^E_1$,  d: $\rho^{LD}$). 
Dashed lines indicate the used edge threshold $e^{p}_{th}=52\,$million\ USD, which was set according to 
the percolation properties of the LD-PIN (a): Above $e^{p}_{th}$, most instances of the LD-PIN disintegrate rapidly.\\
Network connectivity at this value of the edge threshold is believed to contribute dominantly to the \textit{global} properties 
of the LD-PIN, at the same time rendering relative changes comparable over time. The closest coherence between the edge density 
and the spread of financial derivative products is observed around this threshold value, as shown in \hyperref[fig:1]{Fig.\ 1-b},
whereas far away from $e^{p}_{th}$ there are qualitative changes in the evolution of $\rho^{LD}$ (d). The shaded regions in 
panels (b) and (d) indicate the range of the edge threshold $e_{th}=25-55\,$million\ USD within which the non-linear short-term 
memory model (\ref{eq:mm}) achieves a good description for the market values of certain financial derivative products,
such as CDS, ELD and CLD.\\
The disintegration process of the E-PIN for rising edge thresholds is smoother and one does not observe a universal percolation threshold. 
On the other hand, there is almost no qualitative change in the temporal 
evolution of the algebraic connectivity $\lambda_1^E$ for a large range of values of $e_{th}$: 
between $1$-$100\,$million\ USD, it follows approximately the same patterns, while for values above 
$100$-$200\,$million\ USD there is a sudden jump to values around $0.7$, which can be explained through the removal
of network configurations with a low cut-ratio.
\begin{center}
     \begin{figure}[!ht]
          \label{fig:S4}
          \includegraphics[width=\textwidth]{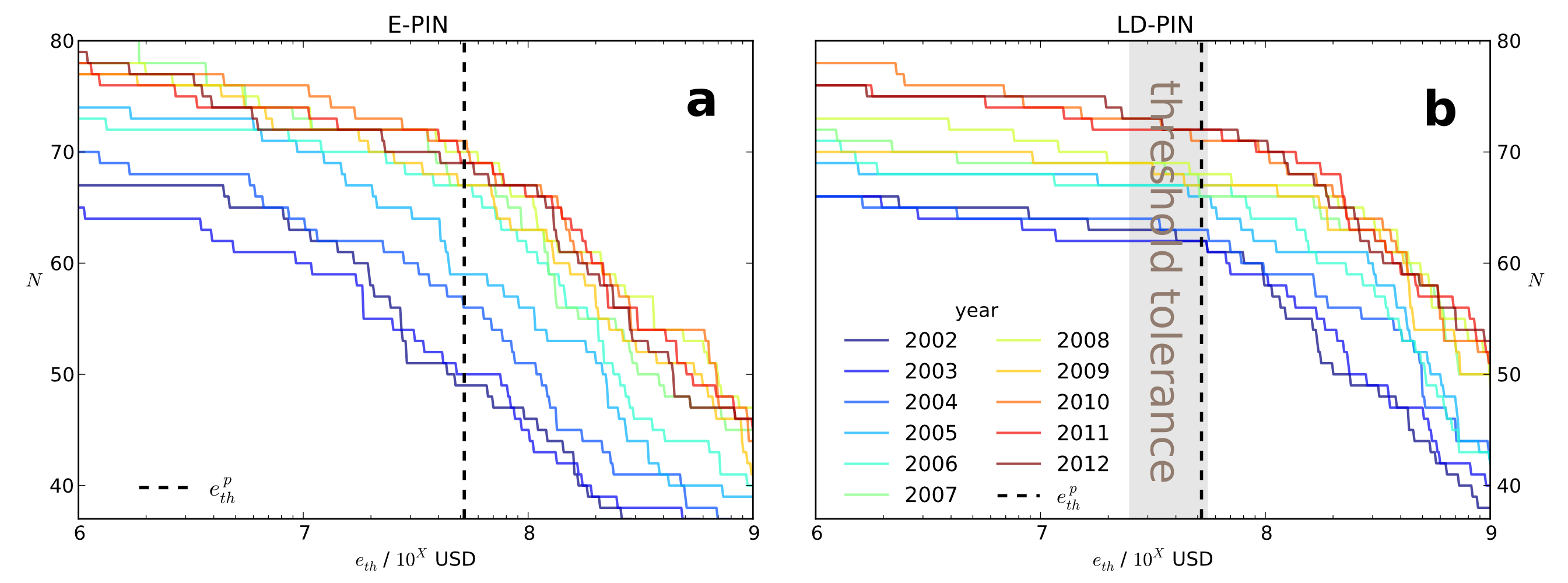}\\
          \includegraphics[width=\textwidth]{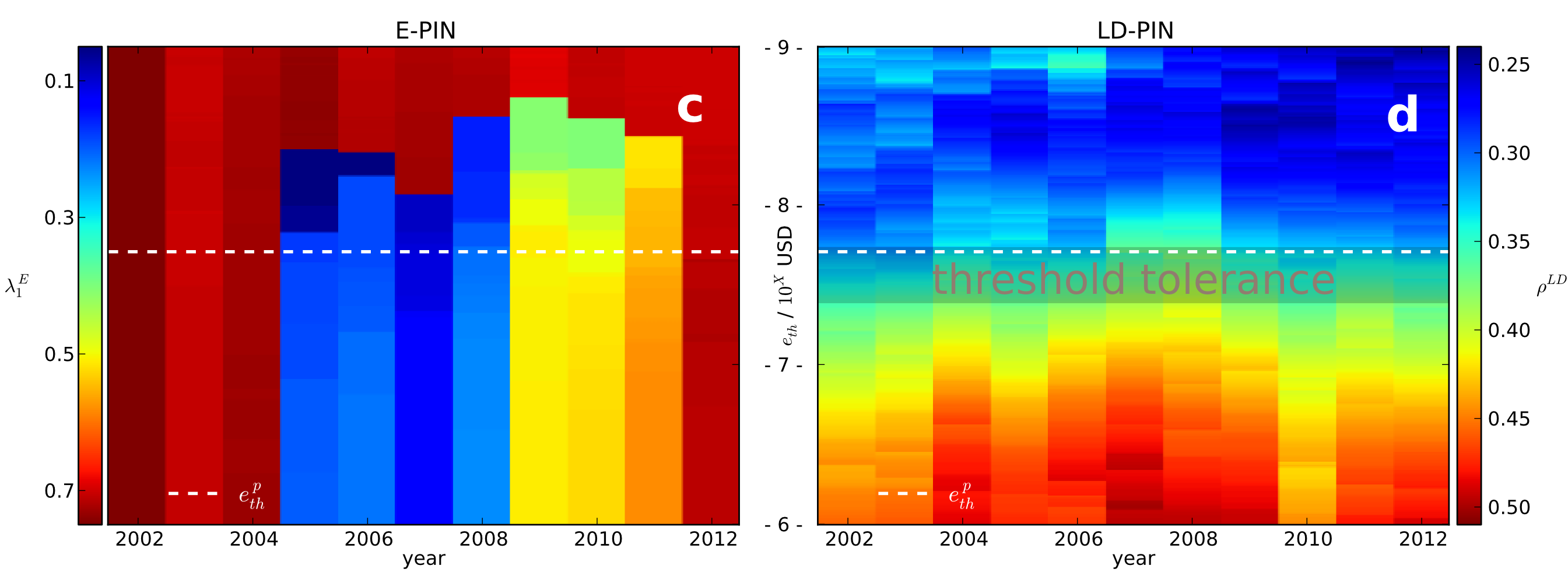}
     \end{figure}
     Figure\ S-4: Threshold dependence of PIN sizes and financial crises indicators.  \textbf{a} and \textbf{c:} E-PIN. \textbf{b} and \textbf{d:} LD-PIN.
\end{center}
\clearpage
\subsection*{S-9: Basic Properties of the SD-PIN}
\begin{figure}[!ht]
          \label{fig:S5}
          \includegraphics[width=1.\textwidth]{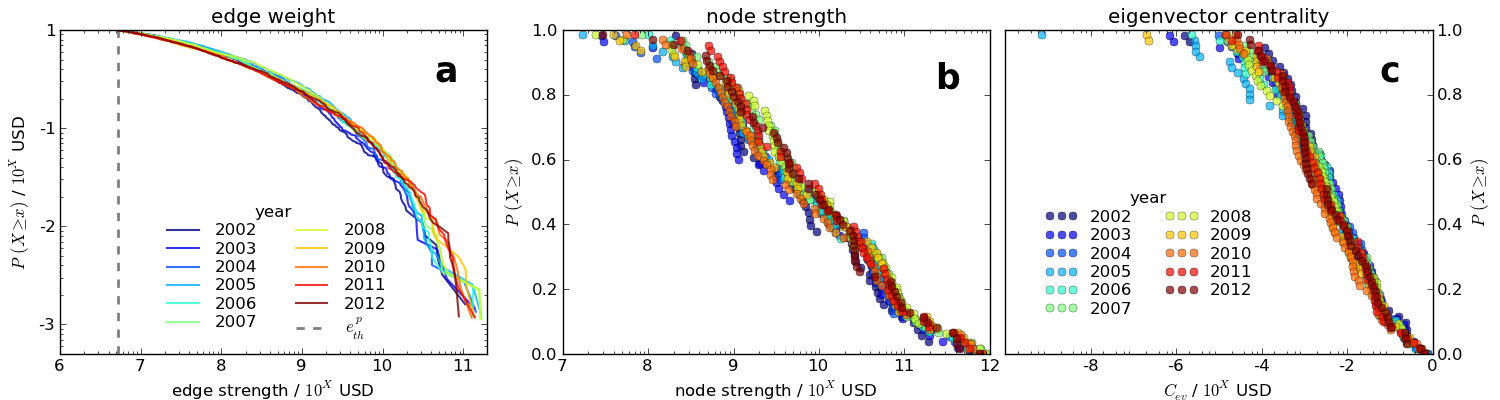}\\[.3cm]
          \includegraphics[width=1.\textwidth]{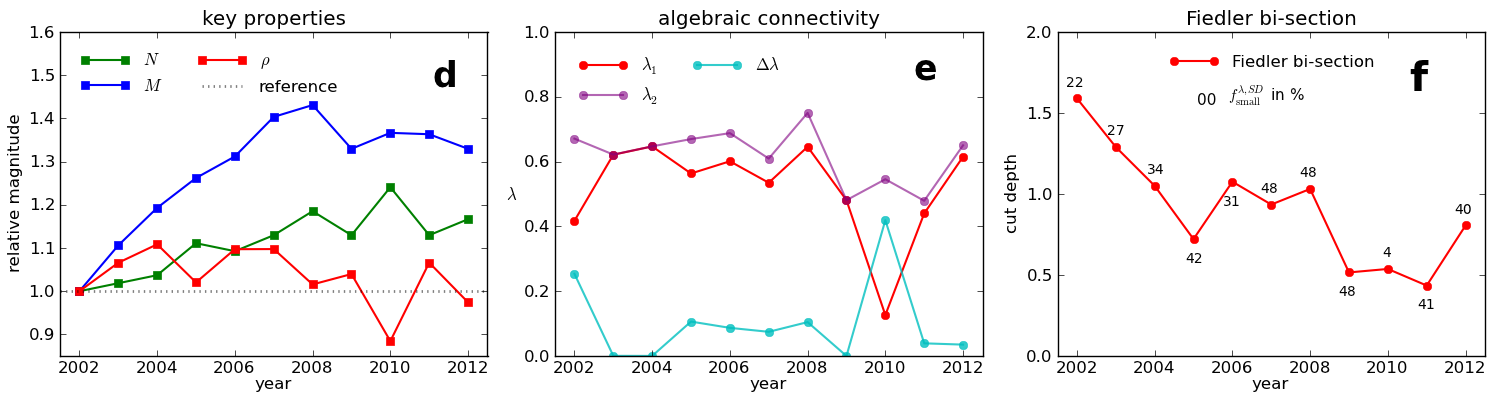}\\
          Figure\ S-5: Basic properties of the SD-PIN. \textbf{a:} Cumulative edge weight distribution. \textbf{b:} Cumulative 
          node strength distribution. \textbf{c:} Cumulative eigenvector centrality distribution. \textbf{d:} Temporal evolution of the number of 
          nodes $N$, the number of edges $M$ and the resulting edge density $\rho^{SD}$ with reverence to their year-2002 values. 
          \textbf{e:} Spectral stability indicators $\lambda_1$, $\lambda_2$ and $\Delta\lambda=\lambda_2-\lambda_1$. \textbf{f:} Cut 
          depth of the Fiedler bi-section, where the numbers indicate the fractions of nodes contained in the smaller 
          sections $s^{\lambda,LD}_{\mathrm{small}}$.
     \end{figure}

\addcontentsline{toc}{subsection}{S-9: Basic Properties of the SD-PIN}
\label{si:9}
In the short-term debt securities (SD) PIN, edges are aggregated cross-border positions of debt instruments which are payable on demand or with a maturity
of up to one year, including  treasury bills, negotiable certificates of deposit, commercial paper, and bankers' acceptances \cite{cpis,bpm6}. The SD-PIN is generated in the same way as the other PIN, namely by extracting the largest strongly-connected component of a threshold graph (see \hyperref[sec:meth]{Methodology}). Because the total volume of SD-securities and the size of positions is generally one order of magnitude smaller than those of E- and LD-investments (see \hyperref[tab:S2]{Tab.\ S-2}), the edge threshold here has been set to $10\,\%$ of that of the E- and LD-PIN, or $e_{th}=5.5\,$million USD. Basic properties of the SD-PIN are summarised in \hyperref[tab:S2]{Tab.\ S-2} and \hyperref[fig:S5]{Fig.\ S-5}. Besides its smaller volume, the biggest difference with respect to the other PIN is the SD-PIN's lower edge density, which is about $\rho^{SD}=0.2$, i.e.\ about $0.1$ lower than those of the E- and LD-PIN. This may, however, be related to the larger number of unavailable data within this statistic.\\
The SD-PIN shows mixed characteristics of the other two PIN. Its edge weight distribution (\hyperref[fig:S5]{Fig.\ S-5-a}) resembles in shape that of the E-PIN (\hyperref[fig:S1]{Fig.\ S-1-a}), while its node strength (\hyperref[fig:S5]{Fig.\ S-5-b}) and eigenvector centrality (\hyperref[fig:S5]{Fig.\ S-5-c}) distributions are similar to those of the LD-PIN (\hyperref[fig:2]{Figs.\ 2-b and\ -d}). The SD-PIN has a pronounced hierarchical structure, as all PIN do, but there is no dominating node, as for the E-PIN where the US is the most central country by far.\\
The temporal evolution of the numbers of nodes $N$ and edges $M$ and the resulting edge density $\rho^{SD}$ are shown in \hyperref[fig:S5]{Fig.\ S-5-d}. The most prominent feature here is the drop in edge density in 2010, which is caused by a considerable increase in the number of nodes. We use the proposed methodology to investigate this phenomenon, looking for specific topological changes at that time. To do so, we consider the Fiedler graph bi-section in combination with the parameter triple $T(SD)$ (see \hyperref[si:6]{App.\ S-6}). The temporal evolution of the algebraic connectivity $\lambda_1^{SD}$ is shown in \hyperref[fig:S5]{Fig.\ S-5-e}, while those of the cut depth $D_{cut}^{\lambda,SD}$ and the fraction of nodes $f^{\lambda,LD}_{\mathrm{small}}$ contained in the smaller sections are given in \hyperref[fig:S5]{Fig.\ S-5-f}. According to $T(SD)$, the drop in edge density was accompanied with the emergence of a sparsely connected and very small group of countries, namely Bahrain, Bermuda and Kuwait. Note that these are exactly the three nodes most-associated with the steep drop of $\lambda_1^E$ of the E-PIN in 2005 (see \hyperref[tab:1]{Tab.\ 1}). This result underpins the findings from the E-PIN and further demonstrates the applicability of the new concepts.\\[.3cm]
\end{document}